\documentclass[aps,pra,twocolumn,superscriptaddress,altaffilletter,secnumarabic,nobalancelastpage,amsmath,amssymb,nofootinbib,longbibliography,floatfix]{revtex4-2}
\usepackage{parskip}
\usepackage{setspace}
\usepackage[version=3]{mhchem} 
\usepackage[utf8]{inputenc}   
\usepackage{CJK}
\usepackage{graphics}   %
\usepackage[caption=false]{subfig}
\usepackage{tikz}
\usepackage{graphicx,float}      %
\usepackage{url}           %
\usepackage{bm}     
\usepackage{subfig}
\usepackage{caption}
\usepackage{mwe}
\usepackage{xcolor}

\usepackage[T1]{fontenc}
\usepackage{amssymb}
\usepackage{amsmath}
\usepackage{amsthm}
\usepackage{amsfonts}
\usepackage{lipsum}
\renewcommand{\vec}[1]{\boldsymbol{#1}}
\usepackage{float}
\usepackage{enumitem}
\usepackage[usestackEOL]{stackengine}

\begin{document}
\linespread{0.75}
\begin{CJK*}{GB}{}
\title{Nuclear cusps and singularities in the non-additive kinetic potential bi-functional from analytical inversion}
\author	   {Mojdeh Banafsheh}
\email          {mbanafsheh@ucmerced.edu}
\affiliation    {Department of Physics, University of California, Merced, Merced, CA 95348, USA}

\author	   {Tomasz A. Wesolowski}
\email 	   {tomasz.wesolowski@unige.ch}
\affiliation{D\'epartement de Chimie Physique 30, Universit\'e de Gen\`eve, Quai Ernest-Ansermet, CH-1211 Gen\`eve 4, Switzerland}
\author        {Tim Gould}
\email	{t.gould@griffith.edu.au}
\affiliation{Queensland Micro- and Nanotechnology Centre, Griffith University, Nathan, QLD 4111, Australia}
\author	   {Leeor Kronik}
\email 	   {leeor.kronik@weizmann.ac.il}
\affiliation 	    {Department of Molecular Chemistry and Materials Science, Weizmann Institute of Science, Rehovoth 76100, Israel}
\author        {David A. Strubbe}
\email	{dstrubbe@ucmerced.edu}
\affiliation    {Department of Physics, University of California, Merced, Merced, CA 95348, USA}

\begin{abstract}
The non-additive kinetic potential $v^{\text{NAD}}$ is a key quantity in density-functional theory (DFT) embedding methods, such as frozen density embedding theory and partition DFT. $v^{\text{NAD}}$ is a bi-functional of electron densities $\rho_{\rm B}$ and $\rho_{\rm tot} = \rho_{\rm A} + \rho_{\rm B}$. It can be evaluated using approximate kinetic-energy functionals, but accurate approximations are challenging.
The behavior of $v^{\text{NAD}}$ in the vicinity of the nuclei has long been questioned, and singularities were seen in some approximate calculations.
In this article, the existence of singularities in $v^{\text{NAD}}$ is analyzed analytically for various choices of $\rho_{\rm B}$ and $\rho_{\rm tot}$, using the nuclear cusp conditions for the density and Kohn-Sham potential. It is shown that no singularities arise from smoothly partitioned ground-state Kohn-Sham densities.
We confirm this result by numerical calculations on diatomic test systems HeHe, HeLi$^+$, and H$_2$, using analytical inversion to obtain a numerically exact $v^{\rm NAD}$ for the local density approximation. We examine features of $v^{\rm NAD}$ which can be used for development and testing of approximations to $v^{\rm NAD}[\rho_{\rm B},\rho_{\rm tot}]$ and kinetic-energy functionals.
\end{abstract}
\maketitle
\end{CJK*}

\renewcommand{\vr}{\textbf{r}}
\newcommand{\vR}{\textbf{R}}
\newcommand{\vs}{v_s}
\newcommand{\vvWr}{v_{\text{vW}'}}
\newcommand{\vvW}{v_{\text{vW}}}
\newcommand{\vbvWr}{\bar{v}_{\text{vW}'}}
\newcommand{\vbvW}{\bar{v}_{\text{vW}}}
\newcommand{\vtNAD}{v_t^{\text{\text{NAD}}}}
\newcommand{\vtvWNAD}{v_{t,\text{vW}}^{\text{\text{NAD}}}}
\newcommand{\vt}{v_t}
\newcommand{\vsNADINV}{v^{\text{NAD/INV}}}
\newcommand{\vtvWNADMOJ}{v_{t}^{\text{\text{NAD}}/\text{vW}}}
\newcommand{\comment}[1]{}
\comment{
\section{Preamble for authors}
TG: We have a lot of similar potentials floating around, and the notation confuses even me,
I propose we settle on the following notation:
\begin{itemize}
    \item $v_s[n](\vr)=-\tfrac{\delta T_s[n]}{\delta n(\vr)}$ is the Kohn-Sham potential
    that yields a density $n$. It obeys $\lim_{|\vr|\to\infty}v_s(\vr)=0$ in nuclear
    potentials.
    \item $\vvWr[n](\vr)=\tfrac12\tfrac{\nabla^2\sqrt{n(\vr)}}{\sqrt{n(\vr)}}$
    is the von Weizs{\"a}cker (vW) potential in square root form. Its asymptotic value
    is a constant in nuclear potentials.
    \item $\vvW[n](\vr)=\tfrac14\tfrac{\nabla^2 n(\vr)}{n(\vr)}
    -\tfrac18|\tfrac{\nabla n(\vr)}{n(\vr)}|^2$ is the vW potential in traditional
    form. In an infinite basis set, $\vvWr=\vvW$ but in a finite basis set
    $\vvWr\neq\vvW$.
    \item $\vbvW[n](\vr)=\vvW[n](\vr)-\lim_{|\vr|\to\infty}\vvW[n](\vr)$
    (and similar for $\vbvWr$) are the vW potentials with the constant (for
    nuclear potentials) removed.
    \item Therefore, in the special case of one orbital, we obtain:
    $\vbvW[n_{1o}]=\vbvWr[n_{1o}]=\vs[n_{1o}]$, because
    $\lim_{|\vr|\to\infty}\vvW(\vr)=-\epsilon_i$. Here,
    $n_{1o}=c|\phi_1|^2$ and $\epsilon_1$ is the eigenvalue of $\phi_1$.
    \item $v_{\rm KS}$ needs to be removed entirely, since it should be identical
    to $v_s$ (debatably up to a constant).
    \item $\vtNAD$ means the general non-additive contribution to the potential. $\vtvWNAD$ means the same quantity with the analytic vW functional.
\end{itemize}
MB: \color{red}{I follow firmly the suggesting of reviewing the notations.
I propose we settle on the following notation which vary slightly from other suggestions:}\color{black}
\begin{itemize}
 \item $\vsNADINV[\rho_{\rm B},\rho_{\rm tot}](\vr)=\frac{1}{2}\frac{\nabla^2\sqrt{\rho_{\rm B}(\vr)}}{\sqrt{\rho_{\rm B}(\vr)}}-\frac{\nabla^2\sqrt{\rho_1(\vr)}}{\sqrt{\rho_1(\vr)}}$, which is exact for when is the case of one-orbital formula and $\int \rho_{\rm B}(\vr)d(\vr)=1$ or $2$.
 \item $\vt[\rho_i](\vr)=-\tfrac{\delta T_s[\rho_i]}{\delta \rho_i(\vr)}$ for any approximation of the $T_s[\rho_i](\vr)$ that is defined via constrained search for any arbitrary $\rho_i(\vr)$. Although it is not used in the text but it's required only for us to understand the following point:
 \item $\vtNAD[\rho_{\rm B},\rho_{\rm tot}](\vr)=$ Eq. \eqref{finvanalyt2}. It confirms the last item of TG  above. Also it is consistent with our previous publication and common literature. 
 \item However I would write it slightly different as:  $\vtvWNADMOJ$ means the same quantity with the analytic vW functional.
 \item Replacing all $n$ by $\rho$ for densities
\end{itemize}
}
\section{Introduction}
When the precise description of large and complex systems is not affordable computationally, they can be partitioned into smaller subsystems to make the calculations feasible. The main quantities of interest often pertain only to a localized region of the whole system. Such a region can be solved separately, with a higher level of theory that is more computationally costly, while the the rest of the system can be solved with computationally cheaper methods \cite{wesolowski2015frozen,sun2016quantum}. %
Examples of this embedding strategy include chromophores in protein environments or aqueous solution \cite{isborn2012electronic}, electrolyte molecules in solvents \cite{barnes2015ab}, organic molecules in aggregates \cite{schulz2019description}, quantum defects in solids \cite{ma2020quantum}, or ions in a plasma with an average-atom model \cite{starett2013electronic}.

Two appealing methods for calculating the electronic structure of complex molecular systems in the framework of density-functional theory (DFT) are frozen-density embedding theory (FDET) \cite{wesolowski1993frozen, wesolowski2008embedding, pernal2009orbital, cortona1991self} and partition DFT \cite{elliott2010partition, nafziger2014density}. 
They allow the total electronic density to be divided into subsystem densities that can be separately calculated, in a formally exact framework.

In calculations based on system-fragmenting methods within the Kohn-Sham DFT framework or QM/MM approaches, the relation between the potential of two subsystems is investigated via the so-called non-additive kinetic potential functional $v^{\text{NAD}}[\rho_{\rm B},\rho_{\rm tot}]$ \cite{Banafsheh}.
This quantity plays a critical role in calculating the correct ground-state density.
In the overlap regions between partitioned densities,  $v^{\text{NAD}}$ takes into account the orthogonality of the wavefunctions of the full system (but not between subsystems \cite{Savin1}).

$v^{\text{NAD}}$ can be evaluated through a kinetic-energy functional \cite{cortona1991self, wesolowski1993frozen}, as a ``decomposable approximation''; the semi-local one most commonly used for $v^{\text{NAD}}$ was introduced in \cite{wesolowski1996accuracy} and tested comprehensively in \cite{wesolowski1997density}. The simple Thomas-Fermi and von Weizs\"{a}cker functionals are found to perform very poorly \cite{Savin1}. In general kinetic-energy functionals \cite{smargiassi1994orbital,huang2010nonlocal} are at a much cruder state of development than exchange-correlation functionals, and perform poorly for $v^{\text{NAD}}$. They can be used for orbital-free DFT \cite{karasiev2014finite} but continue to be an area of active investigation \cite{luo2018simple,shao2021revised,ryczko2022toward}. To go beyond semi-local approximations, $v^{\text{NAD}}$ can be evaluated for real systems through somewhat problematic numerical inversion \cite{de2012exact,Savin1,Savin2009wesolowski,fux2010accurate}, or ``non-decomposable'' approximations specifically for $v^{\rm NAD}$ \cite{lastra2008orbital,polak2022non}.

DFT approximations are evaluated by their capacity to provide well-known properties of the ground-state with high accuracy. 
The cusp relation (or cusp condition) states that for Coulomb potentials, the electron density has a cusp at the position of the nuclei. In DFT-related approaches, the cusp relations \cite{steiner1963charge,liu1995cusp, Kato,nagy2000higher,pack1966cusp} are an important property of an accurately calculated ground-state density, and have corresponding singularities in the Kohn-Sham potential. 
Whether $v^{\rm NAD}[\rho_{\rm B},\rho_{\rm tot}]$ should have such singularities due to nuclear cusps, and whether any approximations produce them, has been unclear. A preliminary investigation with analytic inversion suggested that there were singularities in $v^{\rm NAD}[\rho_{\rm B},\rho_{\rm tot}]$ for diatomic systems \cite{Banafsheh}. However, the following questions remain open: Does $v^{\text{NAD}}[\rho_{\rm B},\rho_{\rm tot}]$ contain singularities at the nuclei for any admissible \cite{Banafsheh} pair of densities $\rho_{\rm B}$ and $\rho_{\rm tot}$ in any Coulomb system? Does $v^{\text{NAD}}[\rho_{\rm B},\rho_{\rm tot}]$ present any other discontinuities? If yes, how are they related to the ground-state charge density?
The presence of singularities is important as a test for $v^{\rm NAD}$ approximations, and potentially to know whether such features pose a numerical challenge in using $v^{\rm NAD}$ for embedding calculations. Correct reproduction of the kinetic-energy density or $v^{\rm NAD}$ features around nuclei can be essential to avoid artificial charge leaks from the nuclei to the environment \cite{lastra2008orbital}, and for calculations of properties involving core electrons, including light elements, X-ray spectra, or warm dense matter in which core orbitals overlap between atoms at high pressure \cite{bekx2020electronic, karasiev2021unraveling}. 

In this article, we theoretically prove the nonexistence at the vicinity of the nuclei of singularities in analytically inverted $v^{\text{NAD}}[\rho_{\rm B},\rho_{\rm tot}]$ from a class of densities that weakly overlap in space, and we show consistent numerical results for model systems.
Section \ref{secTheory} reviews the non-additive potential bi-functional and how it can be constructed from analytical inversion. 
The setup for a specific class of densities for which the inverted potential is free of cusp-like singularities is explained, and we conclude the cases for which singularities at the vicinity of the nuclei are expected.
Section \ref{NC} gives details of how numerical calculations of analytically inverted $v^{\rm NAD}[\rho_{\rm B},\rho_{\rm tot}]$ were carried out for various partitionings of ground-state Kohn-Sham densities. In Sec. \ref{Results} we present results for the diatomic model systems HeHe, HeLi$^+$, and H$_2$ with comparison to calculations from the von Weizs{\"a}cker kinetic functional \cite{weizsacker1935theorie}. We examine the features found in the analytically inverted $v^{\rm NAD}$. Two appendices provide more detailed mathematical analysis of the smooth parts of densities and potentials, and the relation of cusps in densities and singularities in potentials. 

\section{Theory}\label{secTheory}
	We assume finite molecular systems throughout this work. Consider a system described by the DFT Kohn-Sham equations (in atomic units):
	\begin{eqnarray} \label{KS-Eq}
		\bigg[ -\frac{1}{2}\nabla^2 + v_{\rm{KS}} [\rho](\textbf{r}) \bigg] \phi_{i} (\textbf{r}) = \epsilon_{i}\phi_{i} (\textbf{r}),
	\end{eqnarray}
	where $\rho$ is the electronic charge density, $\phi_i$ is a Kohn-Sham orbital, and $\epsilon_i$ is the corresponding Kohn-Sham eigenvalue.
	The Kohn-Sham potential is
 \begin{eqnarray}	
\begin{aligned}
v_{\rm KS}[\rho](\textbf{r}):=v_{\rm ext}(\textbf{r}) + v_{\rm Hxc}[\rho](\textbf{r}),
\end{aligned}
 \end{eqnarray}	
where $v_{\rm Hxc}[\rho](\textbf{r}):=\tfrac{\delta E_{\rm Hxc}}{\delta\rho}$
is the Hartree, exchange and correlation (Hxc) potential obtained
exactly or via an approximation. 

	The density is given by
	 \begin{eqnarray}
\begin{aligned}
\rho({\textbf r}):=\sum_i f_i |\phi_i({\textbf r})|^2
\label{eqn:rhoKS}
\end{aligned}
 \end{eqnarray}
where $f_i$ is the occupation factor of orbital $\phi_i$.
In frozen-density embedding theory \cite{wesolowski2015frozen}, we regard the system as divided into subsystems $j$. The ground-state solution for each is obtained by Kohn-Sham equations for its orbitals $i$:
	\begin{eqnarray}\label{SubSysOrbitals}
	\begin{aligned}
		\bigg[ -\frac{1}{2}\nabla^2 + v_{\rm{KS}} [\rho_j](\textbf{r}) + v_{\rm{emb}} [\rho, \rho_j](\textbf{r}) \bigg] \phi_{ij} (\textbf{r}) \\
		= \epsilon_{ij}\phi_{ij} (\textbf{r})
	\end{aligned}
	\end{eqnarray}
 where the embedding potential is
 	\begin{eqnarray}\label{SubSysEffPot}
 	\begin{aligned}
		 v_{\rm{emb}}[\rho,\rho_j]({\textbf r})= v_{\rm{KS}}[\rho]({\textbf r})-v_{\rm{KS}}[\rho_j]({\textbf r})\\
		 +\frac{\delta T_s[\rho]}{\delta \rho({\textbf r})}-\frac{\delta T_s[\rho_j]}{\delta \rho_j({\textbf r})}.
	\end{aligned}
 	\end{eqnarray}
This formulation relies on the assumption that the potential $v_{\rm KS}[\rho_j]$ exists for each $\rho_j$. In the case of integer $f_i$, this silent assumption is known as the condition of ``non-interacting $v_s$-representability'' \cite{ParrYang1994}. It means that there exists a Kohn-Sham system for which $\rho_j$ is its ground state. This is an admissibility criterion for $\rho_{\rm B}$ and $\rho_{\rm tot}$ in $v^{\text{NAD}}[\rho_{\rm B},\rho_{\rm tot}]$ \cite{Banafsheh}, namely there exist Kohn-Sham systems for which each of them is a ground state.

Evaluation of the differences of Kohn-Sham potentials (external, Hartree, and exchange-correlation) is straightforward. The last two terms at the right-hand side of Eq. \eqref{SubSysEffPot}, based on a kinetic-energy functional $T_s \left[ \rho \right] = -\frac{1}{2} \sum_i f_i \left< \phi_i [\rho] \left| \nabla^2 \right| \phi_i [\rho] \right>$, constitute the non-additive kinetic potential bi-functional $v^{\text{NAD}}[\rho_{\rm B},\rho_{\rm tot}]$. $\phi_i[\rho]$ indicates that the expectation value of the kinetic energy operator is evaluated for optimal orbitals obtained in the constrained search.

The total kinetic energy of a system ($T_s[\rho]$) is the sum over the kinetic energy of all subsystems ($T_s[\rho_j]$) plus an additional ``non-additive'' term ($T^{\rm NAD}[\rho_{\rm A},\rho_{\rm B}]$ in the case of two subsystems) which is due to fermion statistics for electrons and the constrained search definition of the functional $T_s[\rho]$ \cite{gordon_kim}.
The non-additive kinetic potential bi-functional is defined by the pair of densities provided by total ground-state density $\rho_{\rm tot}(\textbf{r})$, and is denoted by $v^{\text{NAD}}[\rho_{\rm B}, \rho_{\rm tot}](\textbf{r})$ where $\rho_{\rm B}(\textbf{r})$ is one of the possible partitions of the total density. $v^{\text{NAD}}$ in fact is the functional derivative of the non-additive kinetic-energy bi-functional:
\begin{eqnarray}
  \begin{aligned}
 v^{\rm NAD}[\rho_{\rm B},\rho_{\rm tot}]({\textbf r}) =\frac{ \delta T_s^{\rm NAD}[\rho,\rho_{\rm tot}](\textbf{r})}{\delta \rho (\textbf{r}) } \Bigg\rvert_{\rho =\rho_{\rm B}}\\
 =\frac{ \delta T_s[\rho_{\rm tot}](\textbf{r})}{\delta \rho_{\rm tot}(\textbf{r})}-\frac{ \delta T_s[\rho_{\rm B}](\textbf{r})}{\delta \rho_{\rm B} (\textbf{r})}
 \end{aligned}
 \label{finvanalyt2}
 \end{eqnarray}
where $\rho_{\rm B}$ and $\rho_{\rm A}=\rho_{\rm tot}-\rho_{\rm B}$ could be partitioned in different ways, as discussed in Sec. \ref{cusptheory}.
We note that an alternate notation convention is used in other work such as Ref. \cite{Banafsheh}, in which the roles of $\rho_{\rm A}$ and $\rho_{\rm B}$ are swapped; i.e. $\rho_{\rm A}$ is the density of the embedded system, and $v^{\rm NAD}[\rho_{\rm A},\rho_{\rm tot}]({\textbf r})$ is the quantity of interest.

The exact form of $\delta T_s[\rho]/\delta \rho$ is not known (except for the von Weizs\"{a}cker formula \cite{weizsacker1935theorie} for the case of one or two electrons, as discussed later), so it needs to be approximated in general.
Explicit semilocal approximations to the kinetic-energy functional \cite{smargiassi1994orbital} $T_s[\rho](\textbf{r})$ in numerical simulations have proven useful for applications such as orbital-free DFT \cite{karasiev2014finite}, but are quite deficient for $v^{\text{NAD}}[\rho_{\rm B},\rho_{\rm tot}]$ \cite{de2012exact, Banafsheh}. Such failures prompted interest in implicit functionals for $v^{\text{NAD}}$, constructed by means of numerical inversion procedures for the Kohn-Sham equation.
Unfortunately,
this numerical inversion is an ill-defined problem if finite basis sets are used, which results in numerical instabilities
and multiple solutions. While approaches have been developed to handle this non-uniqueness \cite{jacob2011unambiguous}, the instabilities remain a problem that plagues Kohn-Sham inversion with finite basis sets \cite{shi2021inverse}. Details of the possible inversion procedures and approximations for construction of $v^{\text{NAD}}$, and their difficulties, were reviewed by Banafsheh and Wesolowski, with numerical examples \cite{Banafsheh}. The Kohn-Sham equations must be inverted twice to obtain $v^{\text{NAD}}[\rho_{\rm B},\rho_{\rm tot}]$ for a given pair of densities, which exacerbates the numerical problems of inversion.
Only for some model systems, and particular partitionings of the total density ensuring that $\rho_{\rm tot} > \rho_{\rm B}$, can $v^{\text{NAD}}[\rho_{\rm B},\rho_{\rm tot}]$ be expressed analytically \cite {Savin1}, and so few results for $v^{\text{NAD}}$ from the exact KS potential have been presented in the literature.
\subsection{One-Orbital Formula} \label{sec:oneorb}

For a Kohn-Sham system described by Eq. \eqref{KS-Eq} with a density as in Eq. \eqref{eqn:rhoKS},
we shall consider the special case in which only one orbital is occupied. In this situation, we are able to analytically invert the Kohn-Sham equation \cite{Banafsheh} and avoid the problems of numerical inversion, a strategy that has been employed in many other studies of the exact Kohn-Sham potential \cite{helbig2009exact}. This is the situation for a system of one electron, or two spin-compensated electrons.

 If this one occupied orbital is real and positive, as is typically the case for the lowest-energy state of a molecule, then $\phi_1({\textbf r})= \sqrt{\rho_1({\textbf r})}$. Eq. \eqref{KS-Eq} can then be rearranged as: 
 \begin{eqnarray}
 \label{KS-Eq-One-Orb}
 v_{\rm KS}(\textbf{r})=\frac{\nabla^2 \phi_1(\textbf{r})}{2\phi_1(\textbf{r})}+\epsilon_1 \\
 \label{KS-Eq-One-rho}
 = \frac{\nabla^2 \sqrt{\rho_1(\textbf{r})}}{2\sqrt{\rho_1(\textbf{r})}}+\epsilon_1
\end{eqnarray}
We define analytical inversion of the density as: 
\begin{eqnarray}\label{InvPotFunct}
v^{\rm inv}[\rho_1](\textbf{r}):= \frac{\nabla^2 \sqrt{\rho_1(\textbf{r})}}{2\sqrt{\rho_1(\textbf{r})}},
\end{eqnarray}
which is numerically equivalent to the von Weizs\"{a}cker formula, Eq. \eqref{VonWei}, as discussed below. Here, $v^{\rm inv}[\rho]$ is the effective potential which reproduces the density $\rho$, which always exists.
(By contrast, the effective potential $v_s[\rho]$, for which $\rho$ is the ground state for the non-interacting electron system, exists only if $\rho$ is $v_s$-representable.)
$v^{\rm inv}[\rho]$ differs from $v_{\rm KS}[\rho]$, which is an approximate potential using $\rho$ as an ingredient.
In the one-orbital formula case, 
	\begin{eqnarray}\label{KSwithInvPotFunct}
					v^{\rm inv}[\rho_1](\textbf{r}) = v_{\rm KS}(\textbf{r})  - \epsilon_1.
  			\end{eqnarray}

	If we multiply both sides of Eq. \eqref{KS-Eq} by $f_1 \phi^*_1({\textbf r})$ we obtain: 
\begin{eqnarray}		
    \begin{aligned}
        	 -\frac{1}{2}\left< \phi_1 \left| \nabla^2 \right| \phi_1 \right> =\epsilon_1|\phi_1({\textbf r})|^2- v_{\rm KS}({\textbf r})|\phi_1({\textbf r})|^2 
   \end{aligned}
   \label{KStoobtainTs}
 \end{eqnarray}	
 Replacing $|\phi_1({\textbf r})|^2$ by $\rho(\textbf{r})$ while identifying the left term in Eq. \eqref{KStoobtainTs} as $T_s[\rho](\textbf{r})$, from a functional derivative we obtain:
\begin{eqnarray}		
    \begin{aligned}
        v_t[\rho](\textbf{r}):=\frac{\delta T_s[\rho]}{\delta\rho(\textbf{r})}=-v_{\rm KS}[\rho](\textbf{r}) + \epsilon_1= -v^{\rm inv}[\rho](\textbf{r})
   \end{aligned}
   \label{eqn:vT}
 \end{eqnarray}	
 
If the exact $T_s[\rho](\textbf{r})$ is known (in Eq. \eqref{finvanalyt2}), then $v^{\rm inv}[\rho](\textbf{r})=-\delta T_s(\textbf{r})/\delta \rho(\textbf{r})$.

\subsection{Cusps, singularities, and the non-additive potential} \label{cusptheory}

\newcommand{\FC}{\mathcal{C}}
\newcommand{\FN}{\mathcal{N}}

Let us consider very generally the relationship between cusps in densities and singularities in potentials -- we shall define both below. Before beginning,
we impose two restrictions on densities and potentials that are assumed throughout
the remainder of this work:
1) densities will be obtained from and yield singularity-free Hxc potentials; 2) the states
considered always have at least one $1s$ orbital at each nucleus, which
dominates the density near each nucleus.
Both restrictions apply to the exact ground-state density and potentials.
They ensure (see conclusions of \cite{march2000cusp}) that cusp conditions~\cite{Kato,bicout1996stochastic,pack1966cusp,nagy2000higher,march2000cusp}
hold for cusps in densities and for singularities in both external potentials and approximated Kohn-Sham
potentials.

Our goal in this section is to explore how cusps in densities manifest as singularities in non-additive potentials.
Since this varies depending on the nature of densities, we first derive a general rule 
and then apply it to examples from the literature and to the work done here.

A nuclear cusp means that the angularly averaged
density obeys
$\lim_{\vr\to\vR_N}|\nabla\rho|=2Z_N\rho$;
a nuclear singularity means that the
potential obeys $\lim_{\vr\to\vR_N}r_Nv\to Z$,
where $\vr_N=\vr-\vR_N$.
We use a short-hand notation to describe
cusps via $e^{-2Z_N|\vr-\vR_N|}$ and singularities
via $-\tfrac{Z_N}{|\vr-\vR_N|}$.
Each cusp and singularity is uniquely described by
$(\vR_N,Z_N)$ for nuclei $N$ in some set,
$N\in \FN$. Sums over $N$ without
extra clarification imply $N\in\FN$.

Note that the notation above addresses behavior near each nucleus, but does not describe every aspect of the system.
The true density and potentials may be written
as
\begin{align} \label{eq:density_cusp}
    \rho(\vr):=&\sum_{N}\rho_{0,N}e^{-2Z_N|\vr-\vR_N|}
    + \rho_{\text{smooth}}(\vr)\;,
    \\ \label{eq:potential_singularity}
    v(\vr):=&-\sum_{N}\frac{Z_N}{|\vr-\vR_N|}
    + v_{\text{non-sing}}(\vr)\;.
\end{align}
where $\rho_{0,N}$ is the density value at $\vR_N$.
Here, $\rho_{\text{smooth}}(\vr)$ has no cusps and
is zero at each nucleus. $v_{\text{non-sing}}(\vr)$
has no singularities, but needs few other restrictions.
Both functions are discussed in some more depth in Appendix~\ref{App:Smooth}.

All subsequent results follow from three theorems below. The densities involved may be ground-state densities, or other densities which have a mapping of the density to the non-interacting potential (Appendix \ref{MoreTwo}).

\emph{\textbf{Theorem 1:}}
The density of any electronic system has a cusp of the form
$\rho(\vr)\approx \rho_{0,N}e^{-2Z_N|\vr-\vR_N|}$ near
every singularity in the external or KS potential, where
$v_{\text{ext}}(\vr)\approx \vs(\vr)\approx -\tfrac{Z_N}{|\vr-\vR_N|}$.

\emph{Proof:} A more general case is a long-known result \cite{Kato,march2000cusp}. Here, we used
that $\rho_{0,N}$ is non-zero, consistent with our second restriction of having a $1s$ orbital, to narrow it down to systems of relevance. Our first restriction extends
it to approximate Kohn-Sham systems.

\emph{\textbf{Theorem 2:}}
If the density of an electronic system has a cusp of the form
$\rho(\vr)\approx \rho_{0,N}e^{-2Z_N|\vr-\vR_N|}$, then the
external and Kohn-Sham potentials have singularities,
$v_{\text{ext}}(\vr)\approx \vs(\vr)\approx
-\tfrac{Z_N}{|\vr-\vR_N|}$.

\emph{Proof:} The result for interacting systems follows from Theorem~1 and the Hohenberg-Kohn theorem \cite{hohenberg1964density,kohn1996density}. The KS result is easily shown for up to two electrons, by using
the von Weizs{\"a}cker potential \cite{weizsacker1935theorie}
 \begin{eqnarray}	
 \label{VonWei0}
v^{\rm vW}[\rho](\textbf{r}):=\frac{\nabla^2\sqrt{\rho (\textbf{r})}}{2\sqrt{\rho (\textbf{r})}} \\
\label{VonWei}
=\frac{\nabla^2 \rho (\textbf{r})}{4\rho (\textbf{r})} -\frac{|\nabla\rho (\textbf{r})|^2}{8\rho^2 (\textbf{r})}\;,
 \end{eqnarray}	
and properties of the Laplacian and gradient. Note that while Eq. \eqref{VonWei} is the standard form of $v^{\rm vW}$, it is analytically equal to the first form in Eq. \eqref{VonWei0} via the identity
\begin{align}
\nabla^2  \sqrt{\rho(\textbf{r})}= \frac{2\rho(\textbf{r})\nabla^2 \rho(\textbf{r})-\left(\nabla \rho(\textbf{r})\right)^2}{4 \left( \rho(\textbf{r})\right)^{3/2}}.
\end{align}
Since $\vs (\textbf{r}) =v^{\rm vW} [\rho](\textbf{r}) +C$, for some constant $C$, the
singularities are inherited by $\vs$. For more
than two electrons one may use the results of
Appendix \ref{MoreTwo}. This extends the known
result for exact potentials~\cite{march2000cusp} to well-behaved approximations consistent with our restrictions.

\emph{\textbf{Theorem 3:}}
There is thus a one-to-one mapping between cusps in the
density and singularities in the external and Kohn-Sham potentials. That is,
\begin{align}
\rho (\textbf{r}) \approx \sum_N\rho_{0,N}e^{-2Z_N|\textbf{r}-\textbf{R}_N|}
\longleftrightarrow
-\sum_N\tfrac{Z_N}{|\textbf{r}-\textbf{R}_N|}\approx v
\end{align}
up to smooth terms. This includes the important special case of no singularities leading to no cusps, which relies on restriction 1 for approximations to DFT.

\emph{Proof:} This follows directly from the previous two
theorems and a recognition that singularities and cusps near
a nucleus at $\vR_N$ are smooth functions near a different nucleus at $\vR_M \neq \vR_N$.

These theorems let us understand how the
non-additive potential in Eq. \eqref{finvanalyt2} behaves
in the vicinity of a nucleus. We use an alternate form $v_{\rm A,B}^{\rm NAD}$ based on $\rho_{\rm A}$ and $\rho_{\rm B}$ here rather than $\rho_{\rm B}$ and $\rho_{\rm tot}$ as in Eq. \eqref{finvanalyt2} to define clearly the nature of the total density. The most general result is that the set of singularities in
\begin{align}
\label{eq:nadd_A_B}
v_{\rm A,B}^{\text{NAD}}[\rho_{\rm A},\rho_{\rm B}]=\vs[\rho_{\rm B}]
- \vs[\rho_{\rm A}+\rho_{\rm B}]
\end{align}
(from Eq.~\eqref{finvanalyt2} and $\vs=-\delta T_s/\delta\rho$)
is equal to the set of singularities from $\rho_{\rm B}$ with subtracted the set of singularities from $\rho_{\rm tot}=\rho_{\rm A}+\rho_{\rm B}$, which follows from Theorem 3.

We note that the functional $v_{\rm A,B}^{\rm NAD}[\rho_{\rm A},\rho_{\rm B}]$ defined in Eq. \eqref{eq:nadd_A_B} resembles the one defined in FDET and analysed in Refs. \cite{lastra2008orbital, polak2022non} but these functionals have different sets of admissible densities.  
In FDET, the non-additive kinetic potential is the functional derivative of $T_s^{\rm NAD}[\rho_{\rm A},\rho_{\rm B}]$ 
with respect to one of the subsystem densities.
For any pair of densities $\rho_{\rm tot}$ and $\rho_{\rm B}$, which are ground states of Kohn-Sham systems, the potentials given in Eqs. \eqref{finvanalyt2} and \eqref{eq:nadd_A_B} are well-defined and equal. The necessary condition that this potential is also equal to the functional derivative of $T_s^{\rm NAD}[\rho_{\rm A},\rho_{\rm B}]$ for a given $\rho_{\rm tot}$ and $\rho_{\rm B}$ is that $T_s^{\rm NAD}[\rho_{\rm A},\rho_{\rm B}]$ exists for $\rho_{\rm A}=\rho_{\rm tot}-\rho_{\rm B}$. It does not exist, however, if $\rho_{\rm tot}(r)-\rho_{\rm B}(r)<0$ on some measurable volume element.

More precisely, if $\rho_{\rm A}$ has a set of cusps
$\FC^{\rm A}:=\{(\vR_N^{\rm A},Z_N^{\rm A})\}_{N\in \FN^{\rm A}}$,
$\rho_{\rm B}$ has a set of cusps
$\FC^{\rm B}:=\{(\vR_N^{\rm B},Z_N^{\rm B})\}_{N\in \FN^{\rm B}}$,
and $\rho_{\rm tot}=\rho_{\rm A}+\rho_{\rm B}$ has a set of cusps
$\FC:=\{(\vR_N,Z_N)\}_{N\in \FN}$,
then the singular part of the non-additive potential in Eq. \eqref{eq:nadd_A_B} is
\begin{align}
    v_{\text{sing}}^{\text{\text{NAD}}}(\vr)=&
    \sum_{N\in\FN^{\rm B}}\frac{-Z_N^{\rm B}}{|\vr-\vR_N^{\rm B}|} +\sum_{N\in\FN}\frac{Z_N}{|\vr-\vR_N|}
\;,
\label{vtSingNAD}
\end{align}
Although we treated $\FC$ as independent above,
it follows from $\rho_{\rm tot}=\rho_{\rm A}+\rho_{\rm B}$ that
$\FC$ can be obtained from
$\FC^{\rm A}$ and $\FC^{\rm B}$ by the following rules:
i) if $\vR=\vR_N^{\rm A}=\vR_M^{\rm B}$ for some $N\in\FN^{\rm A}$
and $M\in\FN^{\rm B}$ then $\FC$ has a combined cusp $(\vR,\tfrac{\rho_{\rm A}(\vR)Z_N^{\rm A}+\rho_{\rm B}(\vR)Z_M^{\rm B}}{\rho_{\rm tot}(\vR)})$;
ii) other cusps in $A$ and $B$ are included
unmodified. 
Either set can be empty (although this would be very
strange for $\FC$), leading to zero for the corresponding sum.

Applying these rules depends on precise details of the embedding or partitioning scheme.
The next sections therefore apply Eq.~\eqref{vtSingNAD} to
the case of smooth partitioning of densities studied here, as well as to some cases from the literature.

\subsubsection{Smooth partitioning of densities}\label{ClassPaiDensi}

The remainder of the manuscript deals with
densities that are partitioned according to
$\rho_{\rm A}(\vr)=w(\vr)\rho_{\rm tot}(\vr)$ and
$\rho_{\rm B}=(1-w(\vr))\rho_{\rm tot}(\vr)$ where $0<w(\vr)<1$
is a smooth, cusp-free and positive function.
$1-w$ therefore has the same qualities as $w$.
In this case, the non-additive potential has no
cusps.

To show this, we recognize that
$\rho_{\rm A}$, $\rho_{\rm B}$ and $\rho_{\rm tot}$ \emph{all} have the same cusps, which follows from the definition of the density,
and from $w$ and $1-w$ being smooth and finite, so
that they only contribute to smooth terms. Therefore,
$\FC^{\rm A}=\FC^{\rm B}=\FC$ and we obtain,
\begin{align}
    v_{\text{sing}}^{\text{NAD,part}}(\vr)=&0
\end{align}

In Sec. \ref{NC}, we numerically apply a smooth cusp-less function to partition the ground-state density of some model diatomic systems of two and four electrons and we show that the corresponding non-additive potential indeed has no singularities at the
nuclei, consistent with theory.

\subsubsection{Embedding with a cusp-free density $\rho_{\rm B}$}
\label{sec:embed}

In some cases, one obtains a density $\rho_{\rm B}$ that is cusp-free in some region of interest, but otherwise has the same cusps as $\rho_{\rm tot}$, as implicitly assumed in Appendix A of Garc\'ia-Lastra \textit{et al.}~\cite{lastra2008orbital}. This situation happens for (e.g.) embedding calculations where some molecules (with cusps at nuclei) are treated at one level of theory and an additional molecule (with existing cusps and new cusps at the additional nuclei) is embedded in the pre-computed set. This is the typical setup for FDET combining different levels of electronic structure theory \cite{wesolowski2015frozen}. Such situations can also arise via a non-smooth partitioning in which $w \left( \mathbf{r} \right)$ has a constant value of 0 in some region. In such a case, the difference between effective potentials $v_s[\rho_{\rm tot}]$ and $v_s[\rho_{\rm B}]$ is not uniquely defined \cite{wesolowski2022}.

As a result, all cusps of $\rho_{\rm B}$ appear in $\rho_{\rm A}$ and $\rho_{\rm tot}$, but
not vice versa. All cusps appear with the same value and at the same nuclear
positions, when they are present, giving $\FC^{\rm A}=\FC$. We use
$\FN_{\rm A\notin B}$ to denote nuclei yielding
cusps in $A$ and $A+B$ that are not in $B$.
It follows that
\begin{align}
    v_{\text{sing}}^{\text{NAD,emb}}(\vr)=&
    \sum_{N\in \FN_{\rm A\notin B}}\frac{Z_N}{|\vr-\vR_N|}\;.
    \label{eqn:vNADemb}
\end{align}

In accordance with this analysis, a nuclear singularity was found in Appendix A of Ref. \cite{lastra2008orbital}. This situation also arises when the density is partitioned not in space but by orbital, so that $\rho_{\rm A}$, $\rho_{\rm B}$, and $\rho_{\rm tot}$ all have the same cusps. An exactly solvable atom-like model system was studied in this way in Ref. \cite{Savin1}, and a nuclear singularity was also found, as expected from our reasoning.

\subsubsection{Use of a finite basis to represent densities}
\label{sec:finitebasis}

Another interesting case is one where densities are obtained using a finite basis set. We first consider
a Slater-type orbital (STO) basis set, which is able to reproduce cusps, but where the resulting cusps are imperfect~\cite{de2012exact}.
In a finite STO basis one obtains
$\rho(\vr\to\vR_N)\approx \rho(\vR_N)e^{-\tilde{Z}_N|\vr-\vR_N|}$ where
$\tilde{Z}_N$ is the finite basis approximation
for $Z_N$. $\tilde{Z}_N\approx Z_N$ varies with choice
of basis, choice of density functional approximation,
and other details of the calculation.

For convenience we assume that all densities
contain all cusps, as in Sec.~\ref{ClassPaiDensi}.
This leads to $\rho_{\rm A}$, $\rho_{\rm B}$ and $\rho_{\rm tot}$
defined by cusp sets
$\FC^{\rm A}=\{(\tilde{Z}_N^{\rm A},\vR_N)\}$,
$\FC^{\rm B}=\{(\tilde{Z}_N^{\rm B},\vR_N)\}$ and
$\FC=\{(\tilde{Z}_N,\vR_N)\}$, respectively.
Importantly, $\vR_N$ is the same in all cases but
$\tilde{Z}_N^{\rm A}\approx \tilde{Z}_N^{\rm B}\approx 
\tilde{Z}_N$ are not the same (but are similar)
because of errors introduced by the finite basis.
The singular part of the non-additive potential is therefore
\begin{align}
    v_{\text{sing}}^{\text{NAD,STO}}
    =&\sum_{N\in\FN}\frac{\tilde{Z}_N-\tilde{Z}^{\rm B}_N}{|\vr-\vR_N|}
    \label{vsingSTO}
\end{align}
where the terms $\tilde{Z}_N-\tilde{Z}^{\rm B}_N$
in the numerator are effectively random
artefacts, defined by the basis set and other
computational and methodological choices.
These artefacts also apply to embedding, per Sec.~\ref{sec:embed}. Eq.~\eqref{vsingSTO} then acts
in addition to the ``exact'' cusps from Eq.~\eqref{eqn:vNADemb}.

Gaussian-type orbitals (GTOs), used in many quantum chemistry calculations, cannot reproduce
cusps at all, unlike STOs, as they are analytic
near nuclei. Nevertheless, they have an effective
analogue to Eq.~\eqref{vsingSTO} for small but
finite $r_N$ in the vicinity of a nucleus.

Of greatest relevance to the present work
is that calculations on a finite grid (adapted to the description of nuclear cusps and singularities, as discussed in Sec. \ref{NC}) can
eliminate these errors entirely. This involves
effective use of numerical methods, chosen
such that derived potentials are as consistent
as possible with the routines used to solve
effective Hamiltonians. Especially, one should use
$\tfrac{\nabla^2\sqrt{\rho}}{2 \sqrt{\rho}}$ (Eq. \eqref{VonWei})
rather than the mathematically equivalent $\tfrac{\nabla^2\rho}{4\rho}
-\tfrac{|\nabla\rho|^2}{8\rho^2}$ (Eq. \eqref{VonWei}) when
computing potentials.

\section{Numerical Calculations}   \label{NC}

To confirm the validity of the analyses above in a case of smooth partitioning of densities, we perform numerical calculations with the all-electron DFT package DARSEC \cite{makmal2009fully}, which is designed for high-precision calculations on diatomic molecules, including Kohn-Sham inversions \cite{garrick2020exact}. In DARSEC, the Kohn-Sham equations are solved self-consistently using a high-order finite difference approach \cite{fornberg1988generation,beck2000real}. 
A real-space grid based on prolate-spherical coordinates is used to describe a system with two atomic centers. The grid is dense near the atoms and increasingly sparse farther away. This atom-adapted grid provides precise information at the vicinity of the nuclei to enable exploration of the density and potential at these points. It also enables treatment of the singular Coulomb potential at the nuclei, unlike the usual Cartesian grids used in real-space codes which are not designed for all-electron calculations \cite{chelikowsky1994finite}. Due to the cylindrical symmetry of diatomic molecules, the three-dimensional problem is reduced to a two--dimensional one in DARSEC. 
In this work, the systems are defined within an ellipse with semiminor radius of 15 Bohr, and use a $115 \times 121$ set of grid points for coordinates $\mu$ and $\nu$. DFT calculations were performed using the local density approximation (LDA) \cite{ceperley1980ground,perdew1992accurate}.

We have carefully examined the numerical precision of our calculations, given the difficulty of describing cusps and singularities numerically, and the possibility of numerical artefacts being mistaken for cusps or singularities. We find robust results in our tests on stencil size and dividing by denominators close to zero \cite{SM}. Tests show a high degree of mirror symmetry within the region of interest $z \in [-4, +4]$ Bohr, in $v^{\rm NAD}[\rho_{\rm B},\rho_{\rm tot}](\textbf{r})$ or $v^{\rm NAD}[\rho_{\rm A},\rho_{\rm tot}](\textbf{r})$ for homonuclear diatomic systems, which is enabled by the symmetry of our weighting function. As seen below, nuclear cusps for the density and singularities for $v_{\rm KS}$ are well reproduced. The only hint of a singularity in $v^{\rm NAD}$ came from using a clearly inadequate stencil size of 2, or an extremely sharp cutoff in $w \left( \vr \right)$ for partitioning, clearly numerical in both cases \cite{SM}. For all calculations in this article, the finite-difference stencil size was set to 12, as per standard recommendations for DARSEC. We additionally demonstrate below excellent agreement between analytically inverted $v^{\rm inv}[\rho]$ and $v^{\rm KS}[\rho]$, and between $v^{\rm NAD}[\rho_{\rm B},\rho_{\rm tot}]$ from analytical inversion and from the von Weizs\"{a}cker potential.

Based on Eq.\eqref{InvPotFunct}, we implemented in our modified version of DARSEC the equation 
\begin{eqnarray}\label{vNADfunctl}
 v^{\text{NAD/INV}}[\rho_{\rm B}, \rho_{\rm tot}](\textbf{r})=v^{\rm inv}[\rho_{\rm B}](\textbf{r})-v^{\rm inv}[\rho_1](\textbf{r})
\end{eqnarray}
where $\rho_1(\textbf{r}) =2 \left| \phi_1(\textbf{r}) \right|^2$ and $\phi_1(\textbf{r})$ is the lowest-energy orbital.
This analytical inversion is appropriate when the conditions for the one-orbital formula are satisfied (Sec. \ref{sec:oneorb}).

\subsection{Von Weizs{\"a}cker potential}

Because the von Weizs{\"a}cker potential $v^{\rm vW}$ is mathematically equivalent to $v^{\rm inv}$ for densities based on one orbital, we can also obtain correct results for $v^{\text{NAD}}$ when using $v^{\rm vW}$ in place of $v^{\rm inv}$ for calculating subsystems of either one or two electrons, as follows:
\begin{eqnarray}\label{vWvNADfunctl}
 v^{\text{NAD/vW}}[\rho_{\rm B}, \rho_{\rm tot}](\textbf{r})=v^{\rm vW}[\rho_{\rm B}](\textbf{r})-v^{\rm vW}[\rho_{1}](\textbf{r})
\end{eqnarray}
using $v^{\rm vW}[\rho(\textbf{r})]$ as given in Eq. \eqref{VonWei}. While, as argued in Sec. \ref{sec:finitebasis}, such results are expected to give less precise results than the approach with Eq. \eqref{VonWei0} (namely $v^{\text{NAD/INV}}$, Eq. \eqref{vNADfunctl}), we demonstrate good agreement between the two formulations which proves the numerical precision of our calculations.
It is important to note the difference between this equation and Eq. \eqref{finvanalyt2}: the second term uses $\rho_1$ not $\rho_{\rm tot}$, as we would have if we were to take von Weizs{\"a}cker as simply an approximation to $\delta T_s/\delta \rho$, as was considered in Ref. \cite{Banafsheh}. Such a formula would be exact only when both $\rho_{\rm B}$ and $\rho_{\rm tot}$ are one or two electrons, i.e. in the case of H$_2$ but not of HeHe or HeLi$^+$. However, by using instead a two-electron density from the orbital $\phi_1$ we can have a correct formula also for HeHe and HeLi$^+$. Such a formula is equivalent, up to a constant, to using $v_{\rm KS}$ instead of $v^{\rm vW}[\rho_{1}]$; we tested this option as well and found very close agreement, consistently with Fig. \ref{Inv_vs_KS} below.

We compare our $v^{\text{NAD}}$ with $v^{\text{NAD/vW}}$ as a benchmark to help examine whether there are any artefacts due to numerical differences from evaluation using the Laplacian or the gradient. 
In Sec. \ref{Results}, we show the very close agreement of $v^{\text{NAD/vW}}$ and $v^{\text{NAD}}$ for our three test systems, demonstrating the numerical precision of our calculations.
 \subsection{Localization of one or two electrons}
 
   For partitioning the ground-state density numerically, we use a smooth distribution function  $0 \leq F(z)\leq 1$ that has no cusps and respects the smoothness of the function explained in Sec. \ref{ClassPaiDensi}. Specifically we use the Fermi-Dirac distribution function
   \begin{eqnarray}\label{Fermi-Dirac}
   F(z-z_0)= \frac{1}{e^{\alpha(z-z_0)}+1}	
   \end{eqnarray}
   where $z_0$ is the cutoff that sets the $z$ at which $F=0.5$, and $\alpha$ is the curve-smoothing parameter. Other similar sigmoid functions could also be used for this purpose.
    By partitioning the total density of the diatomic system aligned along the $z$-axis into two sub-densities we obtain: $\rho_{\rm B}(\textbf{r})= F(z-z_0) \rho_{\rm tot}(\textbf{r})$ and $\rho_{\rm A}(\textbf{r})  =  \rho_{\rm tot}(\textbf{r}) - \rho_{\rm B}(\textbf{r})$.

In a diatomic system of $N=2+2M$ (for integer $M$) electrons where two spin-compensated electrons can be localized around one nucleus, we choose $z_0$ to satisfy the following condition:
  \begin{eqnarray}\label{Omega}
\int \rho_{\rm B}(\textbf{r}) d{\textbf r} = \int  F(z-z_0)\rho_{\rm tot}({\textbf r}) d{\textbf r} = 2 
  \end{eqnarray}
For the case of one-electron localization, $z_0$ can be chosen so that the integral of Eq. \eqref{Omega} is instead $1$.
We find $z_0$ via a binary-search algorithm, with tolerance $10^{-15}$ for the difference of the integral from 2 (or 1). For homonuclear systems, $z_0$ should be exactly zero by symmetry.
We quantify the density overlap in these systems as
\begin{eqnarray} \label{densityoverlap}
\int \rho_{\rm A}(\textbf{r}) \rho_{\rm B}(\textbf{r})d\textbf{r}.
\end{eqnarray}

In the present calculations $\alpha = 20$ Bohr$^{-1}$ was chosen after testing different values \cite{SM}. Too small a value does not constitute localization on one nucleus, and too large a value (for a given grid) leads to numerical discontinuities and artefacts at $z_0$. Unlike the grid spacing, $\alpha$ is not a numerical parameter to be converged, but rather defines the way which we choose to partition the density. The main effect of changing $\alpha$ within a range $15-50$ Bohr$^{-1}$ is that the plateau of $v^{\rm NAD}$ becomes narrower and taller with larger $\alpha$ (a smaller transition region between 0 and 1). This behavior gives an indication of how the width of the transition in other sigmoidal functions would affect $v^{\rm NAD}$. Such effects are simply a feature that must be taken into account consistently with the density partitioned using a given partitioning function.
  
\section{Results and Discussion}\label{Results}

Two classes of diatomic model systems were chosen for this study. One class contains two electrons in total, and one electron was localized around one nucleus; we use the homonuclear H$_2$. In the second class, there are four electrons in total, and two electrons were localized around one nucleus; we use the homonuclear HeHe and the heteronuclear HeLi$^+$. Key properties of these systems for our studies are summarized in Table \ref{tab:summary}.
For each system, we compare the analytically inverted potential with $v^{\text{NAD/vW}}$ from Eq. \eqref{vWvNADfunctl}.

We study systems with weakly overlapping subsystem densities, in part to enable direct comparison with Ref. \cite{Banafsheh}, in which singularities were previously reported in $v^{\rm NAD}$. All calculations are done for a stretched interatomic distance of 6 Bohr, centered on $z=0$. For homonuclear systems, the maximum density overlap between the sub-densities occurs at $z=0$ by symmetry, while for the heteronuclear systems the location of maximum overlap occurs closer to the nucleus with smaller atomic number.

The graphical representations of the results are provided in 1D and 2D.
The 1D plot is the contour along the minimum value of $\mu$ used, which is closest to the interatomic $z$-axis, since there are no grid points at $x=0, y=0$.

\begin{table}
\begin{tabular}{r r r r r r}

\Longstack{\\System} & \Longstack{$N_e$\\total} & \Longstack{$N_e$\\in $\rho_{\rm B}$} & \Longstack{Cutoff\\$z_0$ (Bohr)} & \Longstack{Density overlap\\($e^2/{\rm Bohr}^3$)} \\
\hline
\hline
HeHe & 4 & 2 & 0 & $3.81 \times 10^{-6}$ \\
\hline
HeLi$^{+}$ & 4 & 2 & $-0.29$ & $2.51\times10^{-3}$ \\
\hline
H$_2$ & 2 & 1 & 0 & $1.75 \times 10^{-5}$ \\
\hline
\end{tabular}
\caption{Summary of key properties for diatomic systems used in this work. The density overlap is defined in Eq. \eqref{densityoverlap}.}
\label{tab:summary}
\end{table}

\subsection{Two-electron localization}

\subsubsection{Homonuclear model system: HeHe}

\begin{figure}[h]
\hspace*{-0.5cm} 
\centering
\includegraphics[width=9cm]{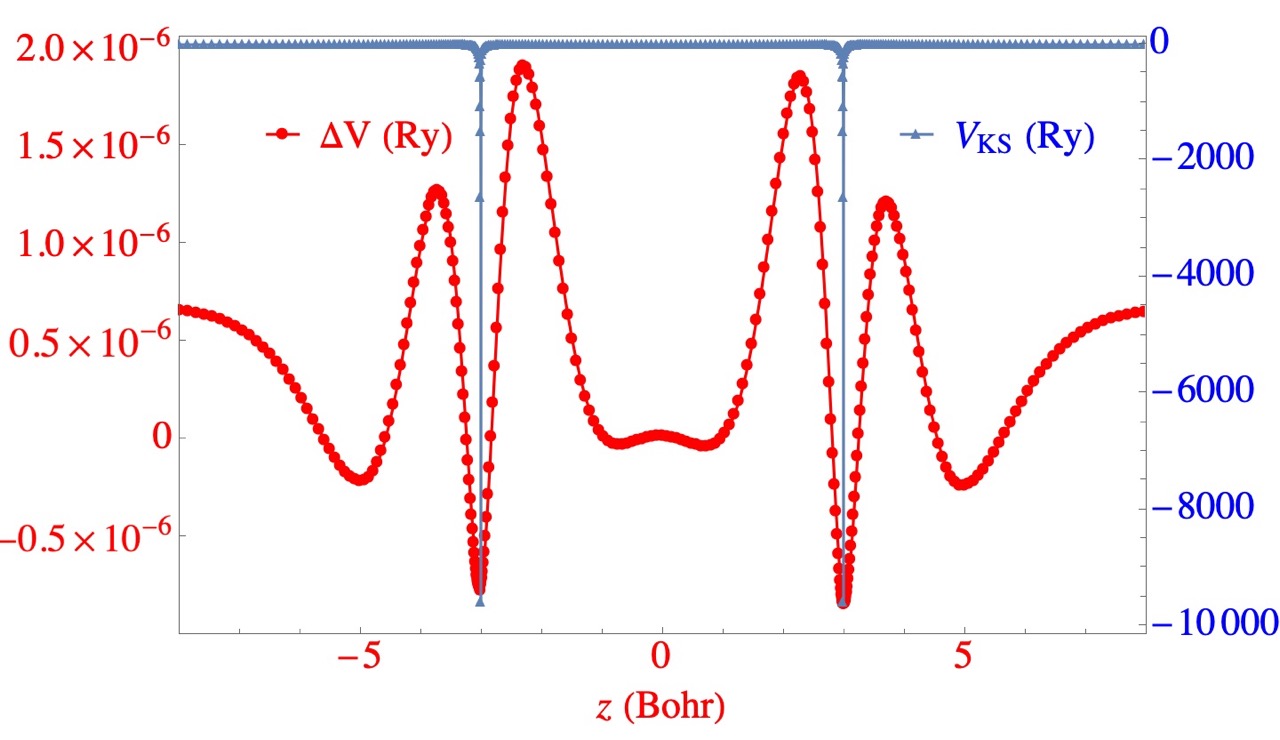}
\caption{Difference between the analytically inverted $v^{\rm inv}[\rho_1](\textbf{r})$ and $v^{\rm KS}(\textbf{r})$ for HeHe, showing agreement to 1 part in $10^{10}$. 
 Blue: $v^{\rm KS}(\textbf{r})$. Red: $\Delta v= v^{\rm KS}(\textbf{r})-v^{\rm inv}[\rho_1](\textbf{r})$.}
\label{Inv_vs_KS}
\end{figure}

We begin the study of this system by demonstrating the numerical precision of our analytical inversion from Eq. \eqref{KS-Eq-One-rho} for $\rho_1= 2|\phi_1({\textbf r})|^2$ and for the eigenvalue $\epsilon_1$, by comparing to the Kohn-Sham potential from the SCF calculation. We find that indeed $\Delta v = v_{\rm KS} -v^{\rm inv}[\rho_1] \approx 0$, as shown in Fig. \ref{Inv_vs_KS}. The difference between the two is approximately 10 orders of magnitude less than the values of the potential, showing excellent precision. The reduction of precision by the square root function is the main contributor to the residual difference. $\Delta v$ is smooth and in particular well-behaved around the nuclei. The singularities at the nuclei are well-reproduced for $v_{\rm KS}$, as per Eq. \eqref{eq:potential_singularity}.

\begin{figure}[h]
\center
\includegraphics[width=8cm]{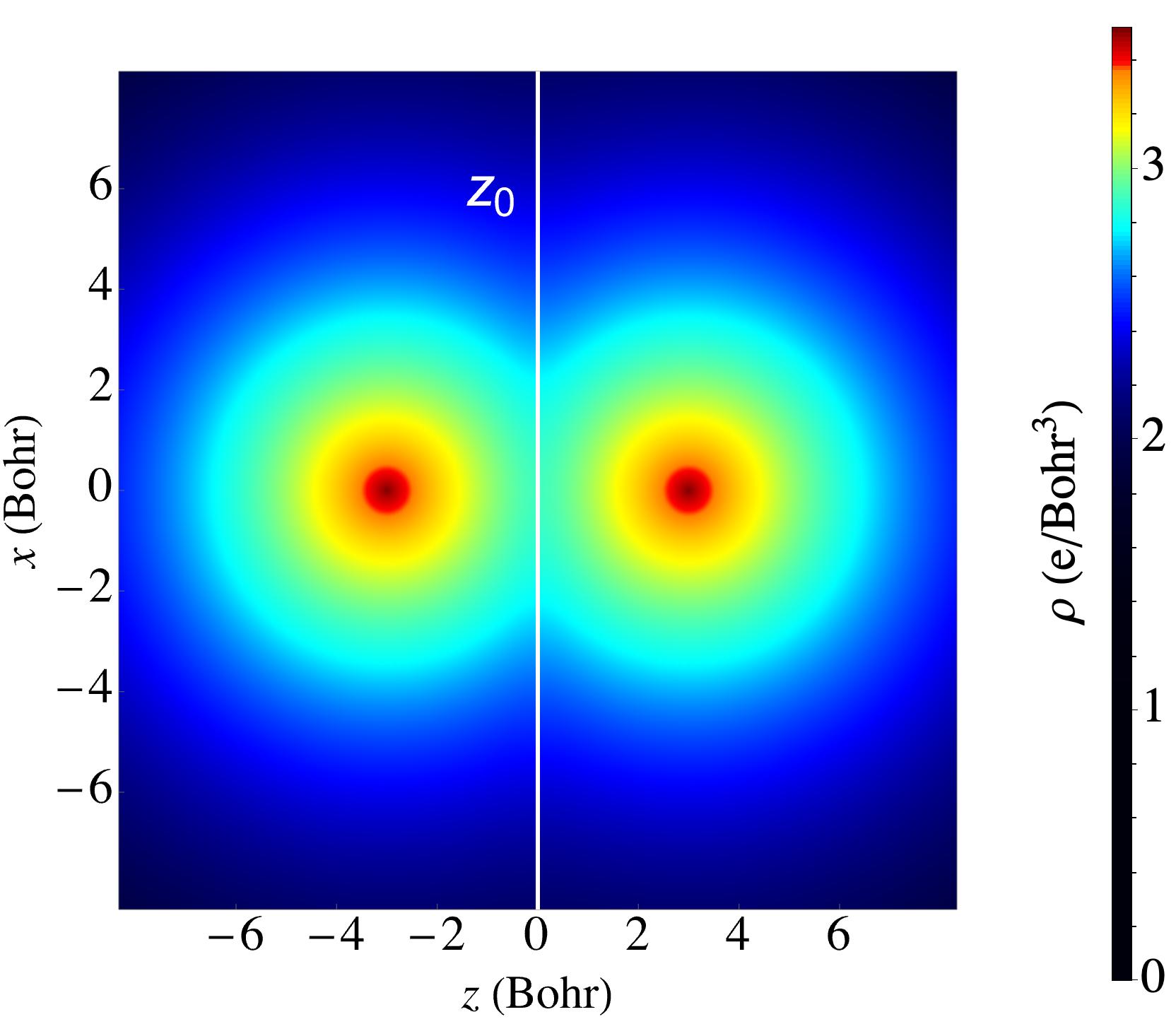}
\caption{Ground-state charge density distribution of HeHe in 2D. The vertical line indicates the cutoff plane where $z=z_0$, dividing the two localized densities which integrate to two electrons. 
}
\label{2dHeHechgDensity}
\end{figure}

\begin{figure}[h]
\begin{tikzpicture}
	\node [anchor=north west] (imgA) at (-0.275\linewidth,.90\linewidth)
			{\includegraphics[width=0.95\linewidth]{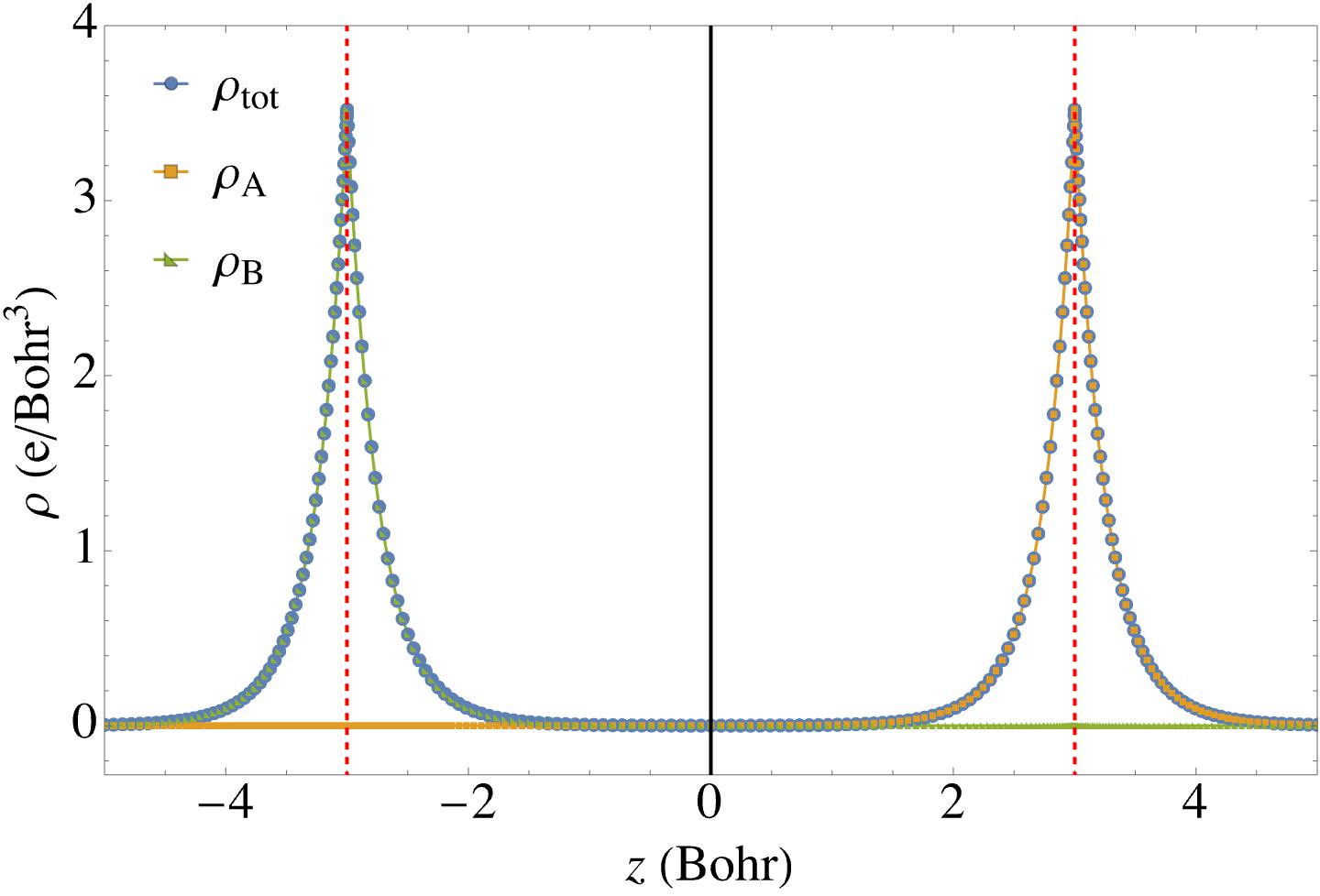}};
   \node [anchor=north west] (imgC) at (-0.275\linewidth,.2\linewidth)
            {\includegraphics[width=1.0\linewidth]{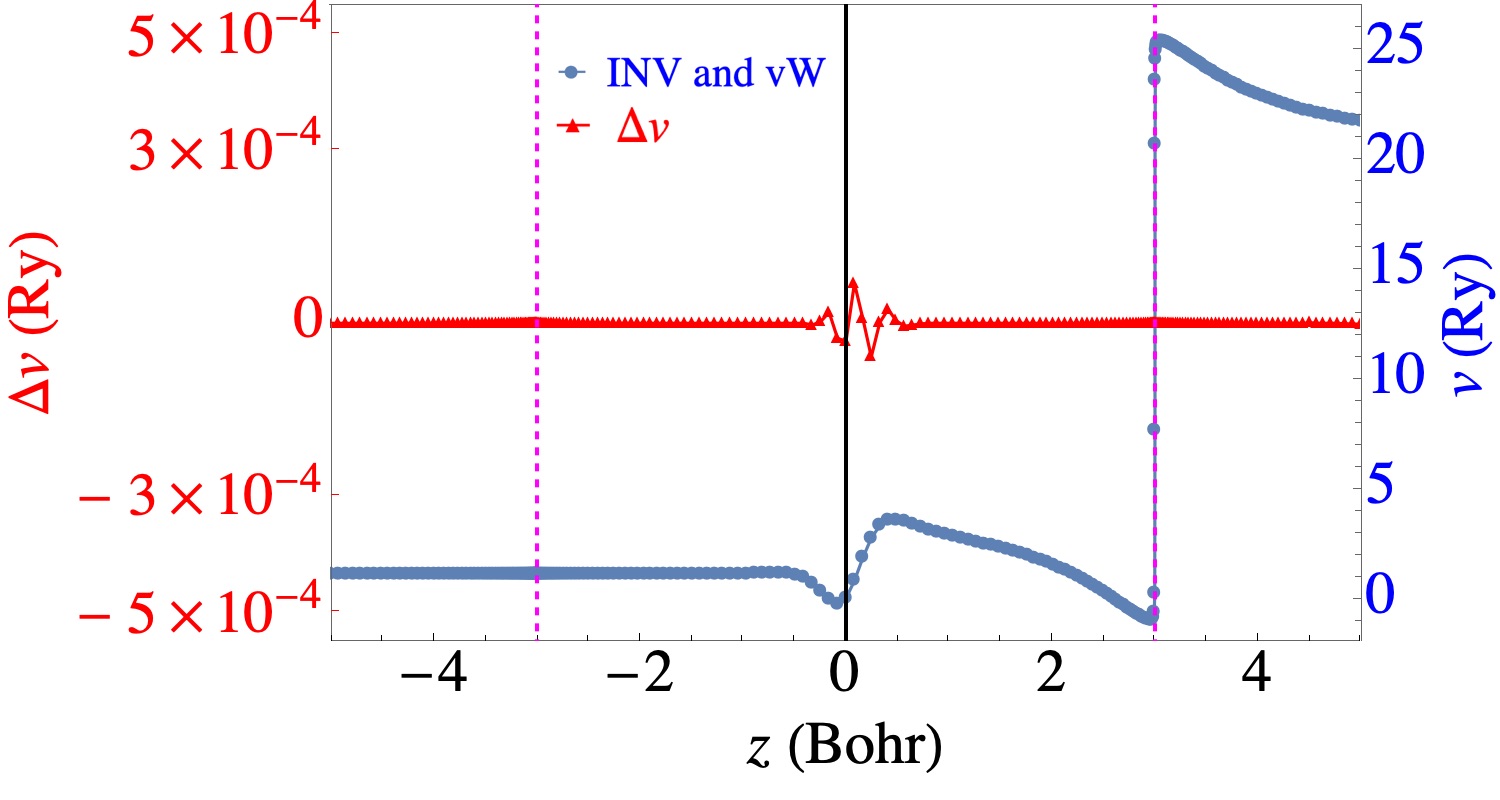}};
    \draw [anchor=north west] (-0.2\linewidth, .95\linewidth) node {\large(a) {\fontfamily{Arial}\selectfont {}}};
    \draw [anchor=north west] (-0.2\linewidth, .26\linewidth) node {\large(b) {\fontfamily{Arial}\selectfont {}}};
\end{tikzpicture}
 \caption{Non-additive kinetic potential for HeHe, in a 1D representation. Purple dashed lines mark the location of the nuclei and a black solid line marks the cutoff $z_0$.%
 (a) Total and partitioned densities. 
 (b) Analytically inverted kinetic potential $v^{\rm NAD/INV}[\rho_{\rm B},\rho_{\rm tot}](\textbf{r})$ (from Eq. \eqref{vNADfunctl}), where the localized density $\rho_{\rm B}$ is on the left, compared to the same from von Weizs{\"a}cker theory (from Eq. \eqref{vWvNADfunctl}). Blue: $v^{\rm NAD/INV}$; Red curve: $\Delta v=v^{\text{NAD/vW}} - v^{\rm NAD/INV}$.
 }
\label{1dHeHechgDensInvvW}
 \end{figure}

The charge density for HeHe including its partitioning is depicted in two dimensions in Fig. \ref{2dHeHechgDensity}. The nuclear cusps are clearly seen, as in Eq. \eqref{eq:density_cusp}. The system is symmetrical about the $z=0$ plane, and therefore the cutoff $z_0$ from Eq. \eqref{Fermi-Dirac} is exactly at $z=0$. %
Consequently, $\rho_{\rm A}$ and $\rho_{\rm B}$ both integrate to 2, each localized around a different nucleus with close to zero charge density in the vicinity of the opposite nucleus, as shown in 1D in Fig. \ref{1dHeHechgDensInvvW}(a). 
\vspace{0.0001mm}
We now proceed to analyze $v^{\rm NAD}$, as shown in 1D in comparison to the input densities in Fig. \ref{1dHeHechgDensInvvW}. For $\rho_{\rm B}$ localized on the left, we see a shape which starts with a small positive value, and has a small attractive well at the overlap region of $z=0$, a barrier, and another small attractive well at the right nucleus. This surprising attractive well at the other nucleus can be attributed to the feature discussed in Sec. \ref{ClassPaiDensi}, in which $\rho_{\rm B}$ must have a cusp at both nuclei; the well induces the extra density for the cusp at the other nucleus. There is a wall at the location of the right nucleus with a plateau afterward, preventing $\rho_{\rm B}$ from entering that region. The wall/plateau is similar to results found with finite basis sets \cite{de2012exact}, though in that case the wall was not exactly at the nucleus but displaced slightly in the direction of the other nucleus. While the wall is a sharp feature, it is smooth with several points along it, and there is no sign of any cusps or singularities. We compare $v^{\rm NAD/INV}$ to $v^{\text{NAD/vW}}$ in Fig. \ref{1dHeHechgDensInvvW}(b) and find excellent agreement with differences around 1 part in $10^{5}$, found mainly in the overlap region, where the division by small $\rho$ is most ill-conditioned.

\begin{figure}[h]
\begin{tikzpicture}
	\node [anchor=north west] (imgA) at (-0.21\linewidth,.90\linewidth)
			{\includegraphics[width=0.95\linewidth]{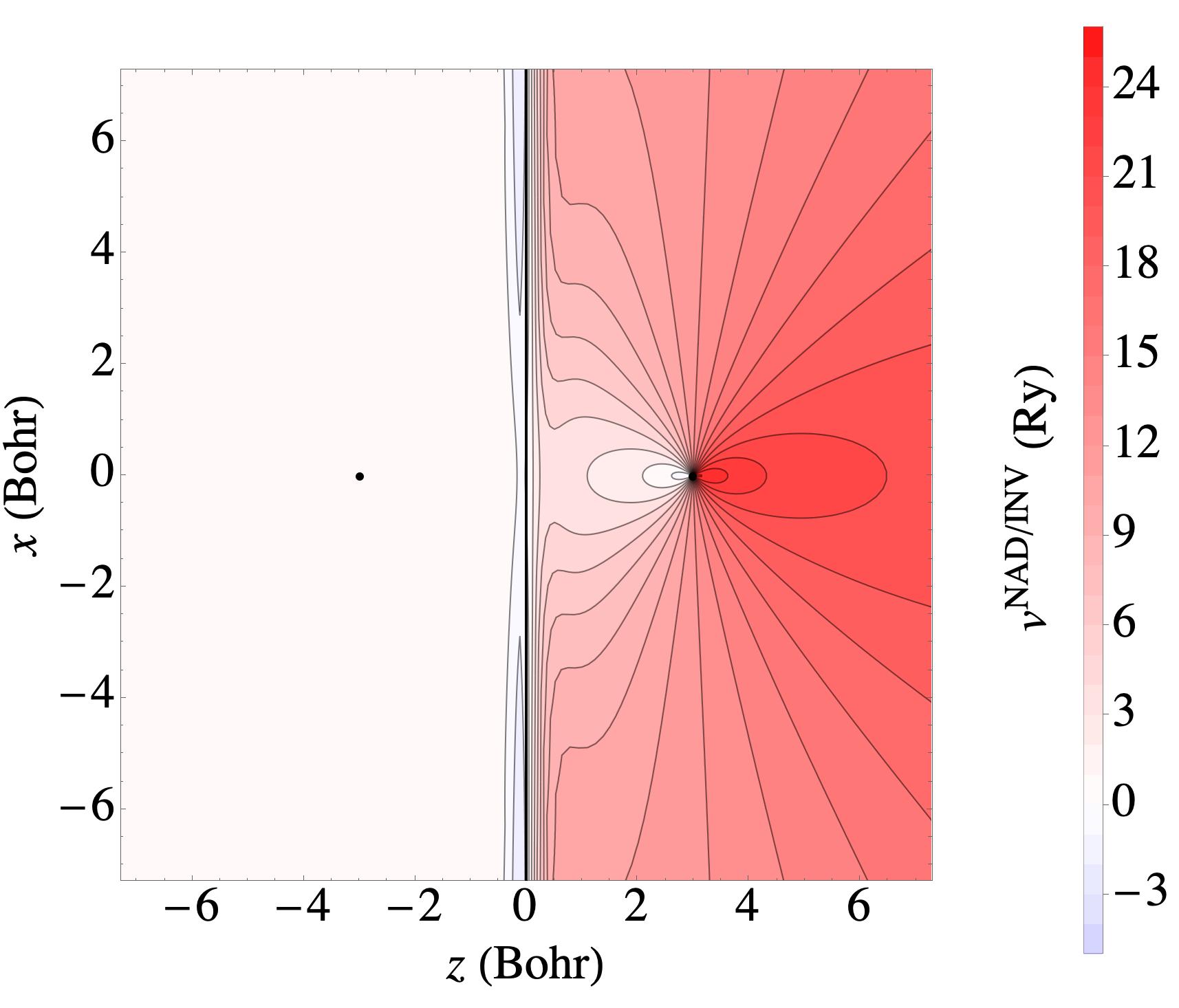}};
   \node [anchor=north west] (imgC) at (-0.21\linewidth,0\linewidth)
            {\includegraphics[width=0.95\linewidth]{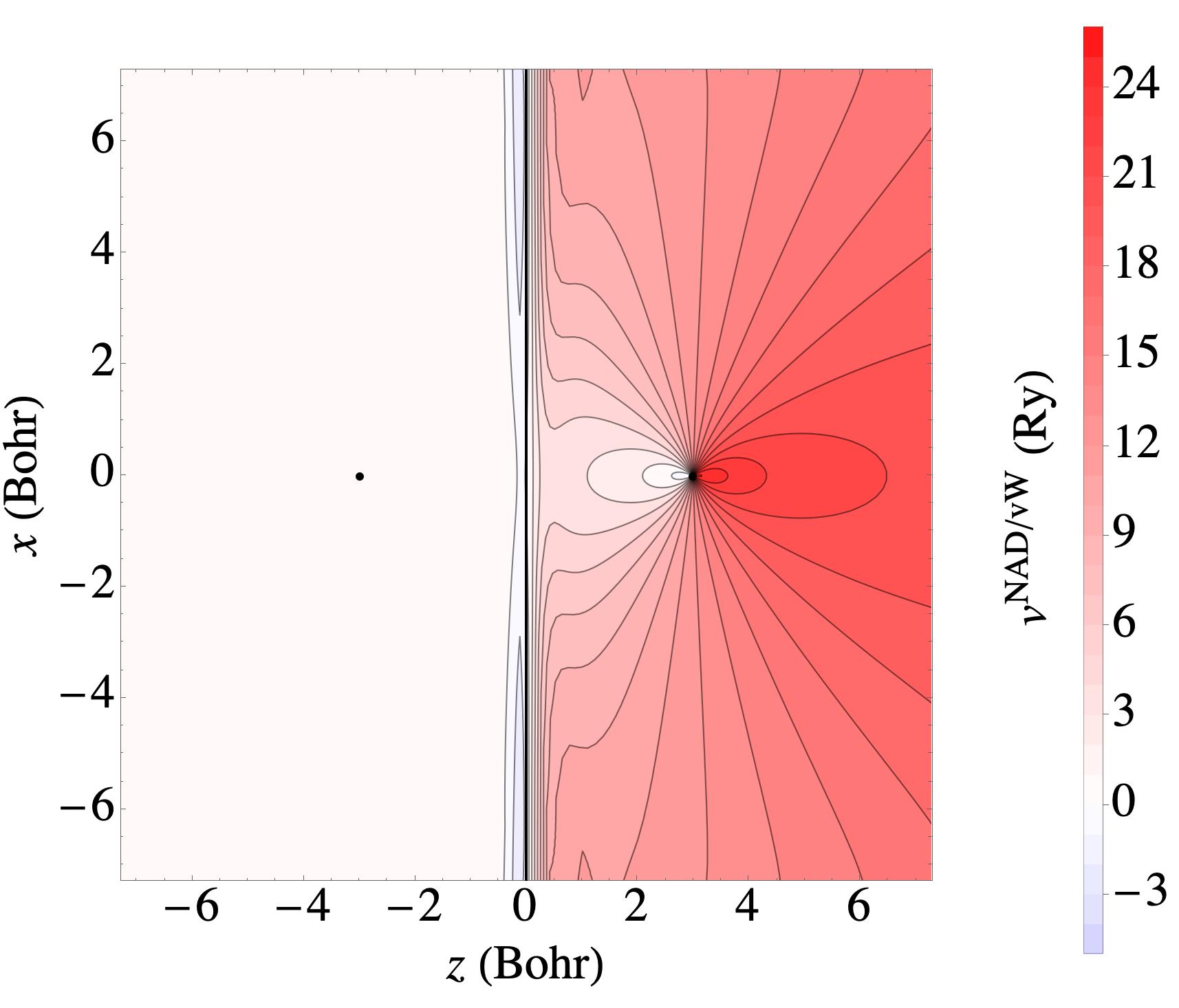}};
\draw [anchor=north west] (-0.1\linewidth, .83\linewidth) node {\large(a) {\fontfamily{Arial}\selectfont {}}};
 \draw [anchor=north west] (-0.1\linewidth, -.07\linewidth) node {\large(b) {\fontfamily{Arial}\selectfont {}}};
                 \end{tikzpicture}
 \caption{Non-additive kinetic potential for HeHe in a 2D representation, comparing (a) $v^{\text{NAD/INV}}$ with (b) $v^{\text{NAD/vW}}$, both in Ry. The density is localized on the left nucleus. Black dots mark the nuclei, and a black line marks the cutoff $z_0$. 
 }  
 \label{2DHeHevW2DHeHe2dVNAD}
 \end{figure}

We further examine $v^{\rm NAD}$ in the 2D representation to see the behavior away from the $z$-axis (Fig. \ref{2DHeHevW2DHeHe2dVNAD}). We see that $v^{\rm NAD}$ is essentially constant in the $z<0$ region. The well around $z=0$ becomes increasingly attractive away from the $z$-axis, but the well at the right nucleus is located only near the $z$-axis. The plateau falls away slowly from the $z$-axis. The steep wall at the right nucleus can be clearly seen, but no cusps or singularities are visible here either. We again compare $v^{\rm NAD/INV}$ and $v^{\text{NAD/vW}}$, now in 2D, and see no perceptible difference except small deviations around $x=6$ Bohr near the cutoff, where the density is very small (Fig. \ref{2DHeHevW2DHeHe2dVNAD}).


\subsubsection{Heteronuclear model system: HeLi$^{+}$}

\begin{figure}[h]
\begin{tikzpicture}
	\node [anchor=north west] (imgA) at (-0.275\linewidth,.90\linewidth)
	{\includegraphics[width=0.95\linewidth]{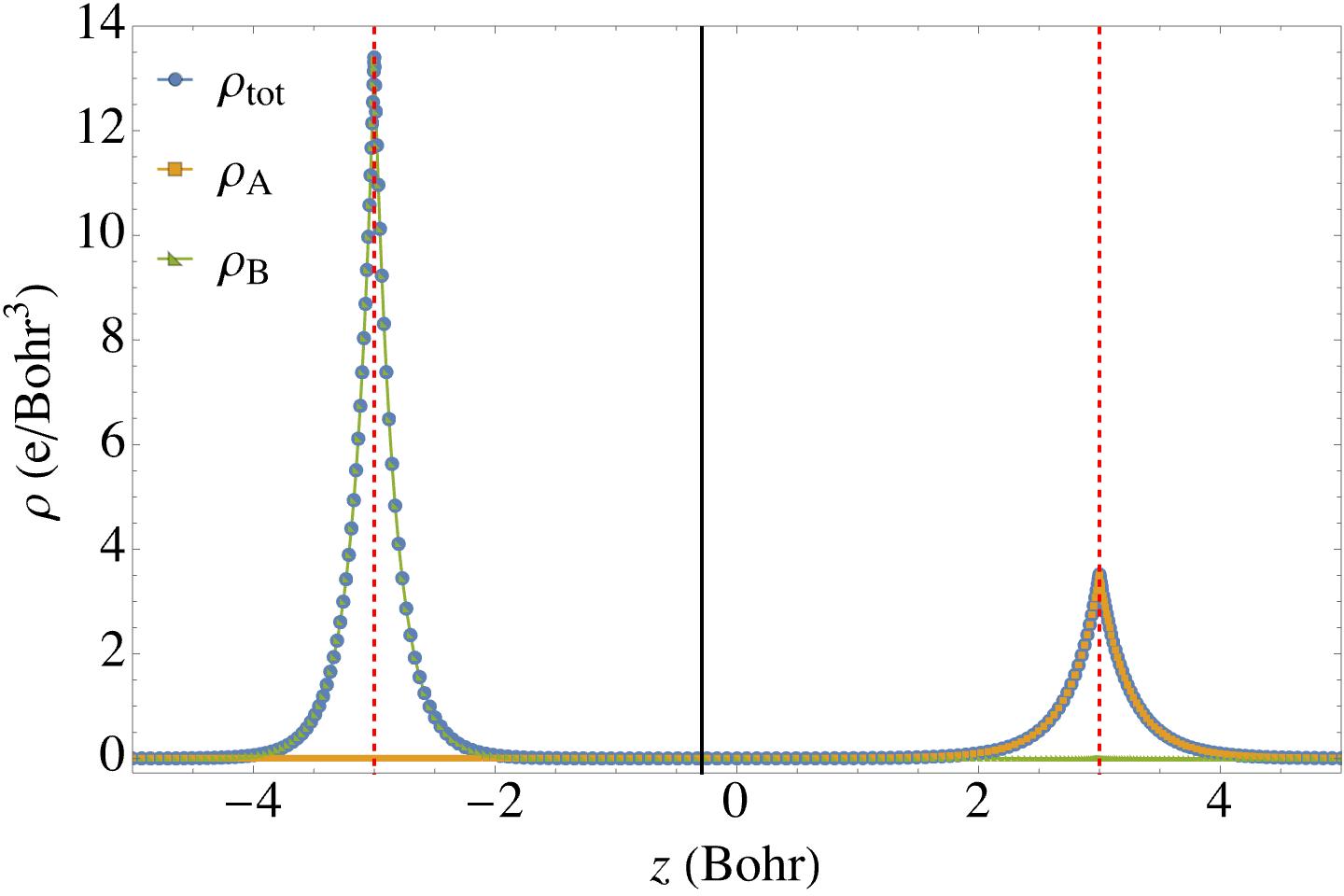}};%
   \node [anchor=north west] (imgC) at (-0.275\linewidth,.18\linewidth)
            {\includegraphics[width=1.0\linewidth]{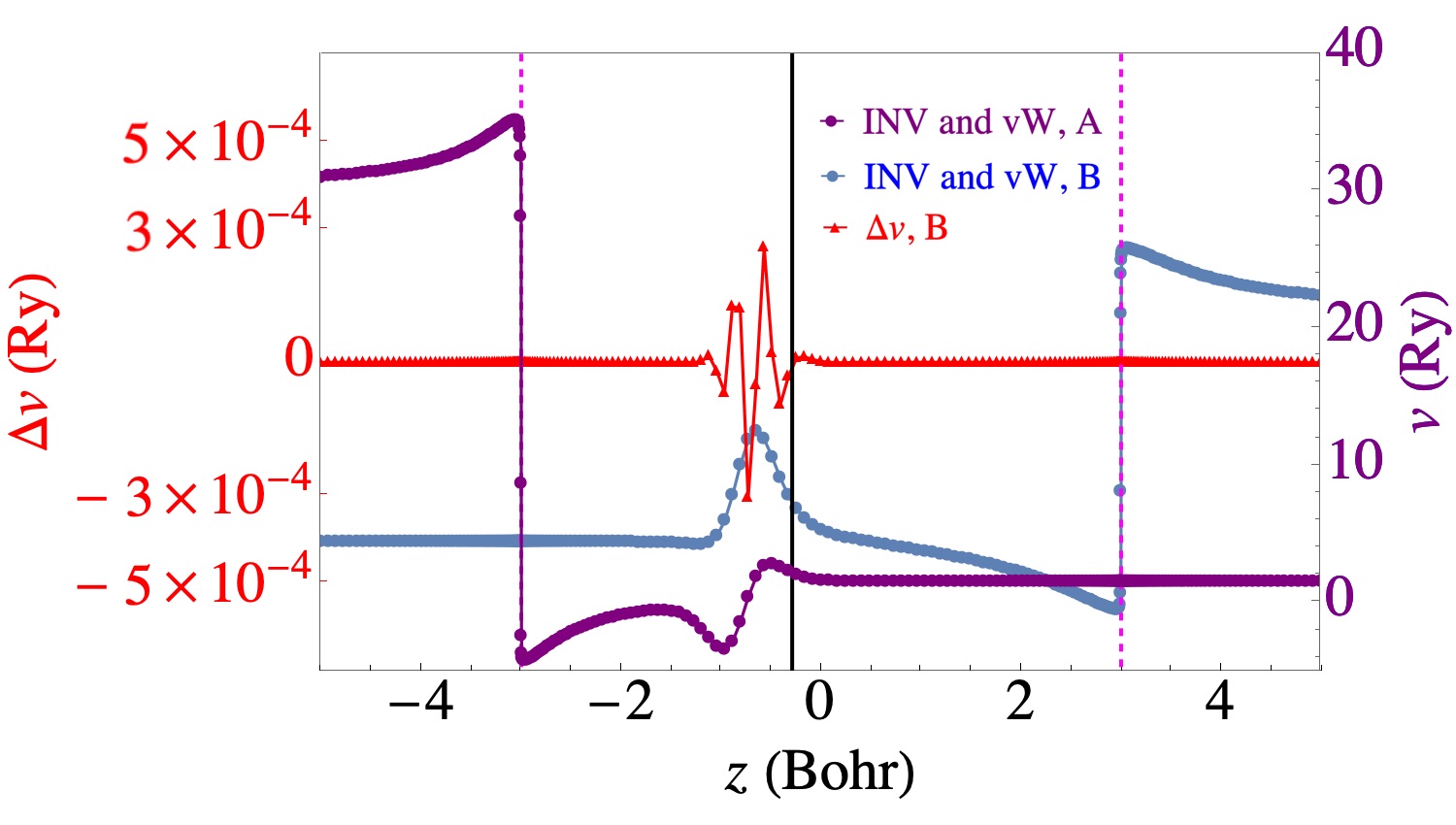}};
    \draw [anchor=north west] (-0.15\linewidth, .95\linewidth) node {\large(a) {\fontfamily{Arial}\selectfont {}}};
    \draw [anchor=north west] (-0.15\linewidth, .22\linewidth) node {\large(b) {\fontfamily{Arial}\selectfont {}}};
\end{tikzpicture}
 \caption{Non-additive kinetic potential for HeLi$^+$ (with Li on the left side), in a 1D representation. Purple dashed lines mark the location of the nuclei and a black solid line marks the cutoff $z_0$.
 (a) Total and partitioned densities, with a taller peak for Li.
 (b) Analytically inverted kinetic potential $v^{\rm NAD/INV}[\rho_{\rm B},\rho_{\rm tot}](\textbf{r})$ (from Eq. \eqref{vNADfunctl}) which is localized on the Li side (blue), compared to the same from von Weizs{\"a}cker theory (from Eq. \eqref{vWvNADfunctl}), with $\Delta v=v^{\text{NAD/vW}} - v^{\rm NAD/INV}$ in red. Also shown is $v^{\rm NAD/INV}[\rho_{\rm A},\rho_{\rm tot}](\textbf{r})$ which is localized on the He side (purple).}
 \label{1dLiDensDelvInvandvw}
 \end{figure}

Next we consider a heteronuclear system, again with 4 electrons: HeLi$^{+}$. We localize 2 electrons on the Li atom side, which has $z < 0$ and is on the left in our plots. We find a cutoff $z_0 = -0.29$ Bohr, which is slightly closer to the nucleus with the larger atomic number, since it has a more steeply decaying density according to Eq. \eqref{eq:density_cusp} (Fig. \ref{1dLiDensDelvInvandvw}(a)). The density overlap is larger than for HeHe (Table \ref{tab:summary}). %

\vspace{1.0mm} 
 
In this case, we can actually calculate two distinct quantities, $v^{\rm NAD/INV}[\rho_{\rm B},\rho_{\rm tot}](\textbf{r})$ (localized on Li) and $v^{\rm NAD/INV}[\rho_{\rm A},\rho_{\rm tot}](\textbf{r})$ (localized on He), as shown in Fig. \ref{1dLiDensDelvInvandvw}(b). For HeHe, the corresponding quantities are identical by symmetry. The shape and magnitude of $v^{\rm NAD}[\rho_{\rm B},\rho_{\rm tot}](\textbf{r})$ is similar to that of HeHe, with a small positive value on the left and a wall and plateau on the right. In the overlap region, however, there is not a step but rather a small peak just to the left of $z_0$, and then a small slightly attractive well just on the left side of the He nucleus, next to the steep wall. This well again can be related to the need to induce a cusp. $v^{\rm NAD}[\rho_{\rm A},\rho_{\rm tot}](\textbf{r})$ has a similar shape, albeit flipped horizontally, and with a smaller peak, two attractive wells, and a higher wall. The fact that there are two attractive wells is in common with He localized in HeHe.

\begin{figure}[h]
\begin{tikzpicture}
	\node [anchor=north west] (imgA) at (-0.21\linewidth,.90\linewidth)
			{\includegraphics[width=0.95\linewidth]{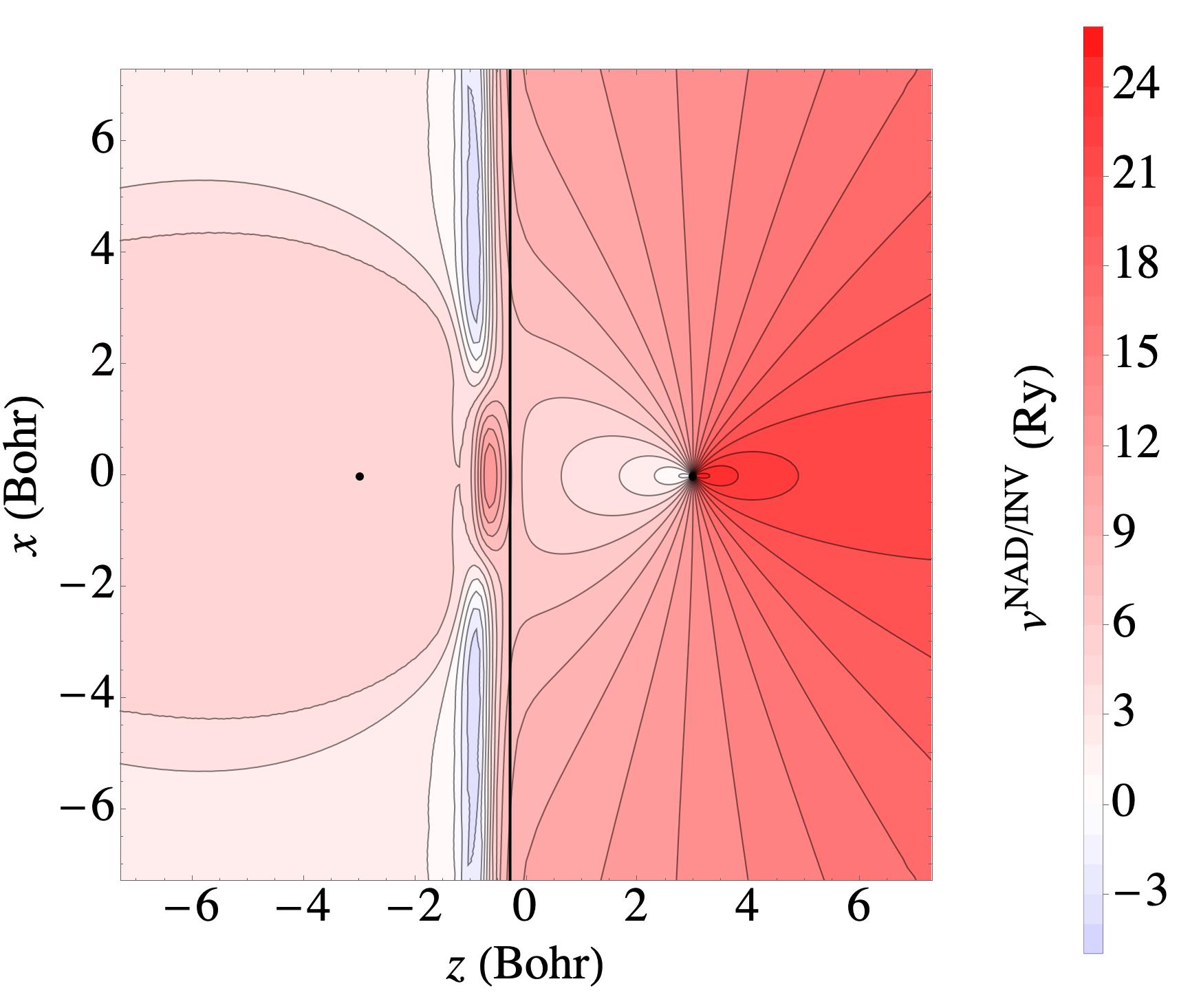}};
   \node [anchor=north west] (imgC) at (-0.21\linewidth,.004\linewidth)
            {\includegraphics[width=0.95\linewidth]{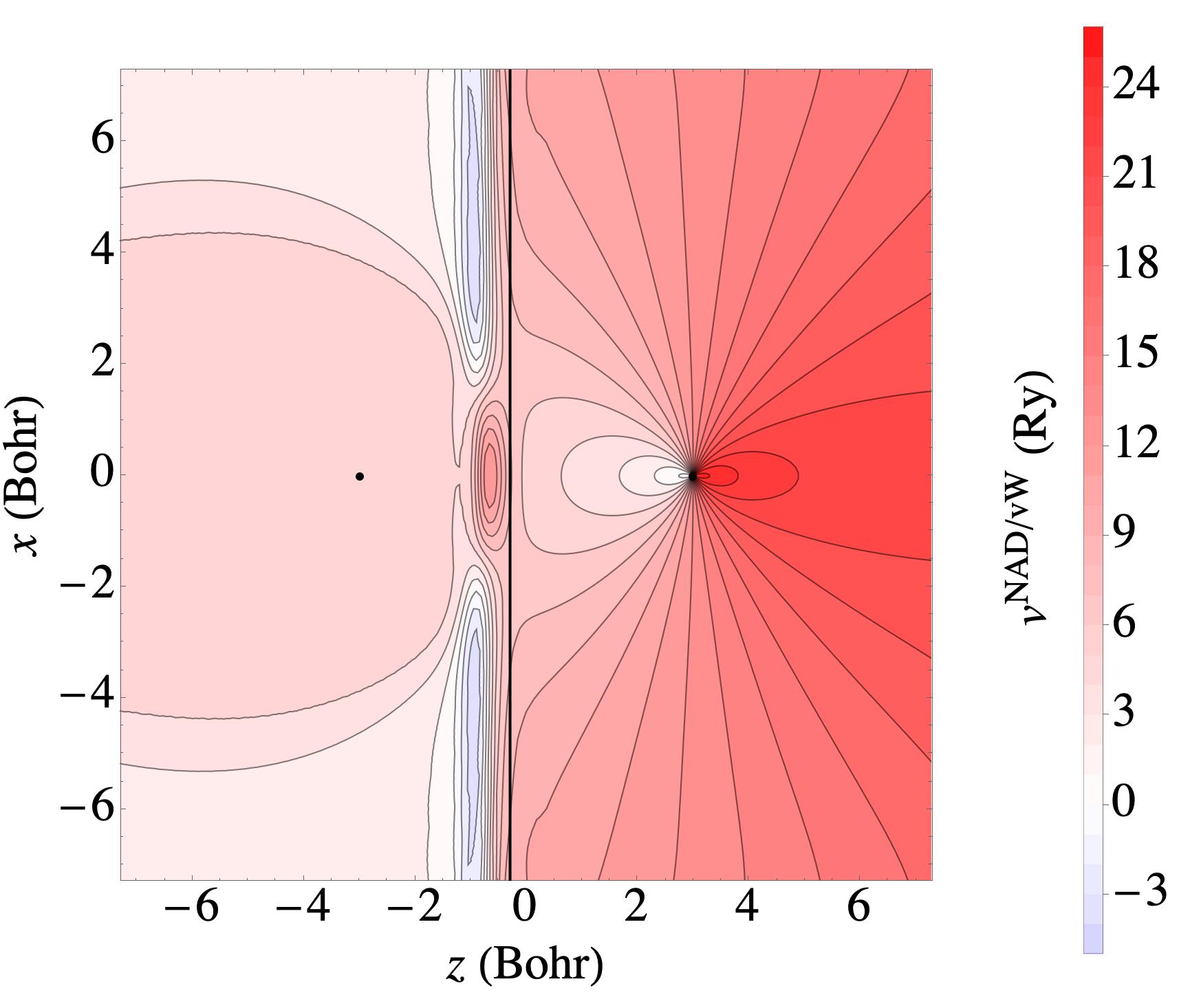}};
    \draw [anchor=north west] (-0.1\linewidth, .83\linewidth) node {\large(a) {\fontfamily{Arial}\selectfont {}}};
    \draw [anchor=north west] (-0.1\linewidth, -.07\linewidth) node {\large(b) {\fontfamily{Arial}\selectfont {}}};
\end{tikzpicture}
 \caption{
 Non-additive kinetic potential for HeLi$^+$ in a 2D representation, comparing (a) $v^{\text{NAD/INV}}[\rho_{\rm B},\rho_{\rm tot}](\textbf{r})$ with (b) $v^{\text{NAD/vW}}[\rho_{\rm B},\rho_{\rm tot}](\textbf{r})$, both in Ry. The density is localized on Li, on the left. Black dots mark the nuclei, and a black line marks the cutoff $z_0$.}
 \label{2DLiHe2dvW2DLiHe2dVNAD}
 \end{figure}

We compare $v^{\rm NAD/INV}[\rho_{\rm B},\rho_{\rm tot}](\textbf{r})$ and $v^{\rm NAD/vW}[\rho_{\rm B},\rho_{\rm tot}](\textbf{r})$ and find again excellent agreement with differences $\Delta v$ around 1 part in $10^4$, largest around the peak next to the overlap region. The 2D plots in Fig. \ref{2DLiHe2dvW2DLiHe2dVNAD} also show no perceptible difference between $v^{\rm NAD/INV}$ and $v^{\rm NAD/vW}$, except small deviations around $x=6$ Bohr near the cutoff. The shapes are similar to HeHe, except that $v^{\rm NAD}$ is not constant on the left, and decreases away from the $z$-axis. The positive peak just to the left of $z=0$ becomes an attractive well like HeHe, centered about 5 Bohr from $z$-axis. No cusps or singularities are seen in the 1D or 2D plots.

\vspace{1.0mm} 

\subsection{One-electron localization for H$_2$}

\begin{figure}[h]
\begin{tikzpicture}
	\node [anchor=north west] (imgA) at (-0.275\linewidth,.90\linewidth)
			{\includegraphics[width=0.95\linewidth]{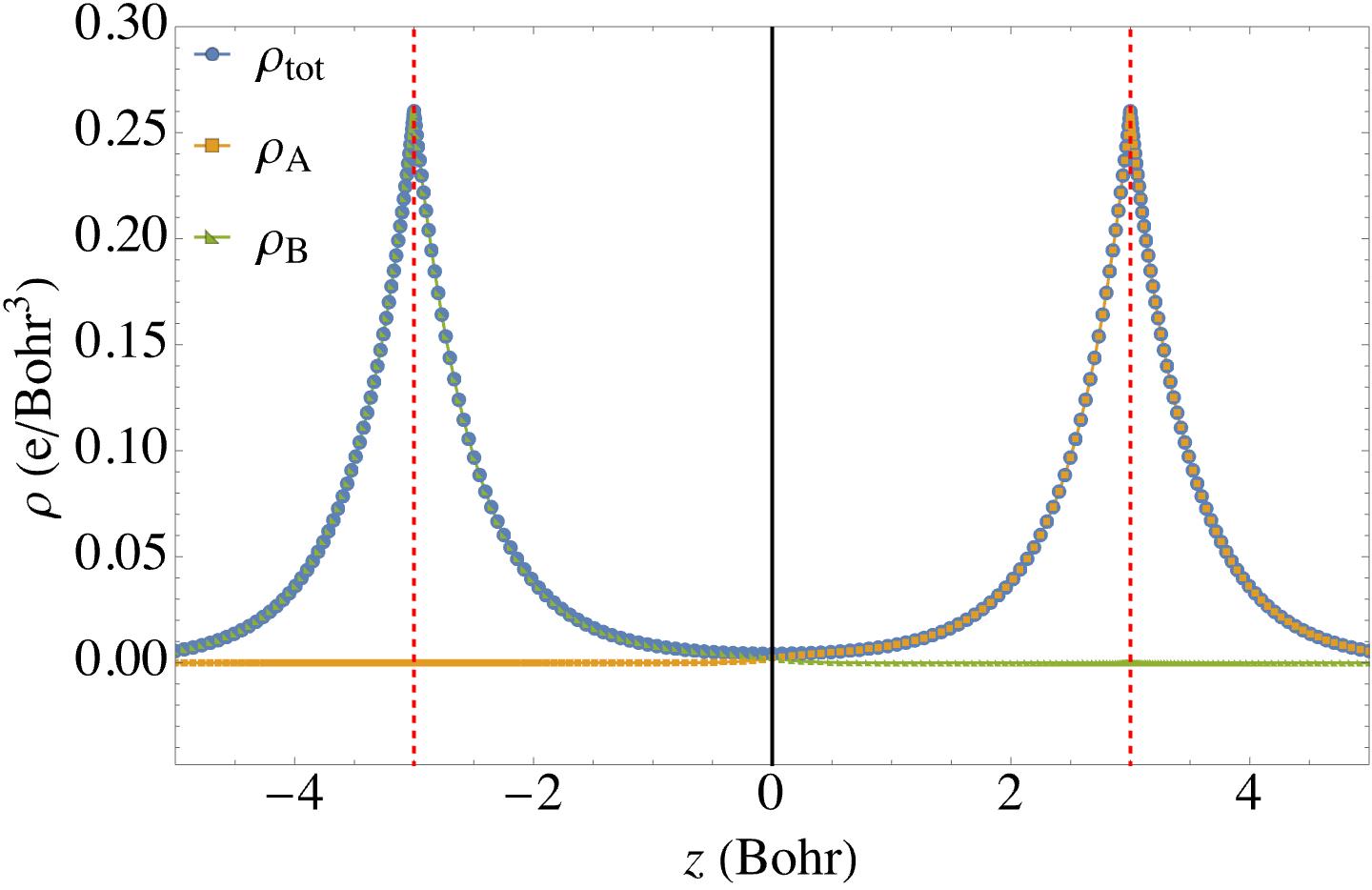}};	%
   \node [anchor=north west] (imgC) at (-0.275\linewidth,.18\linewidth)
            {\includegraphics[width=1.0\linewidth]{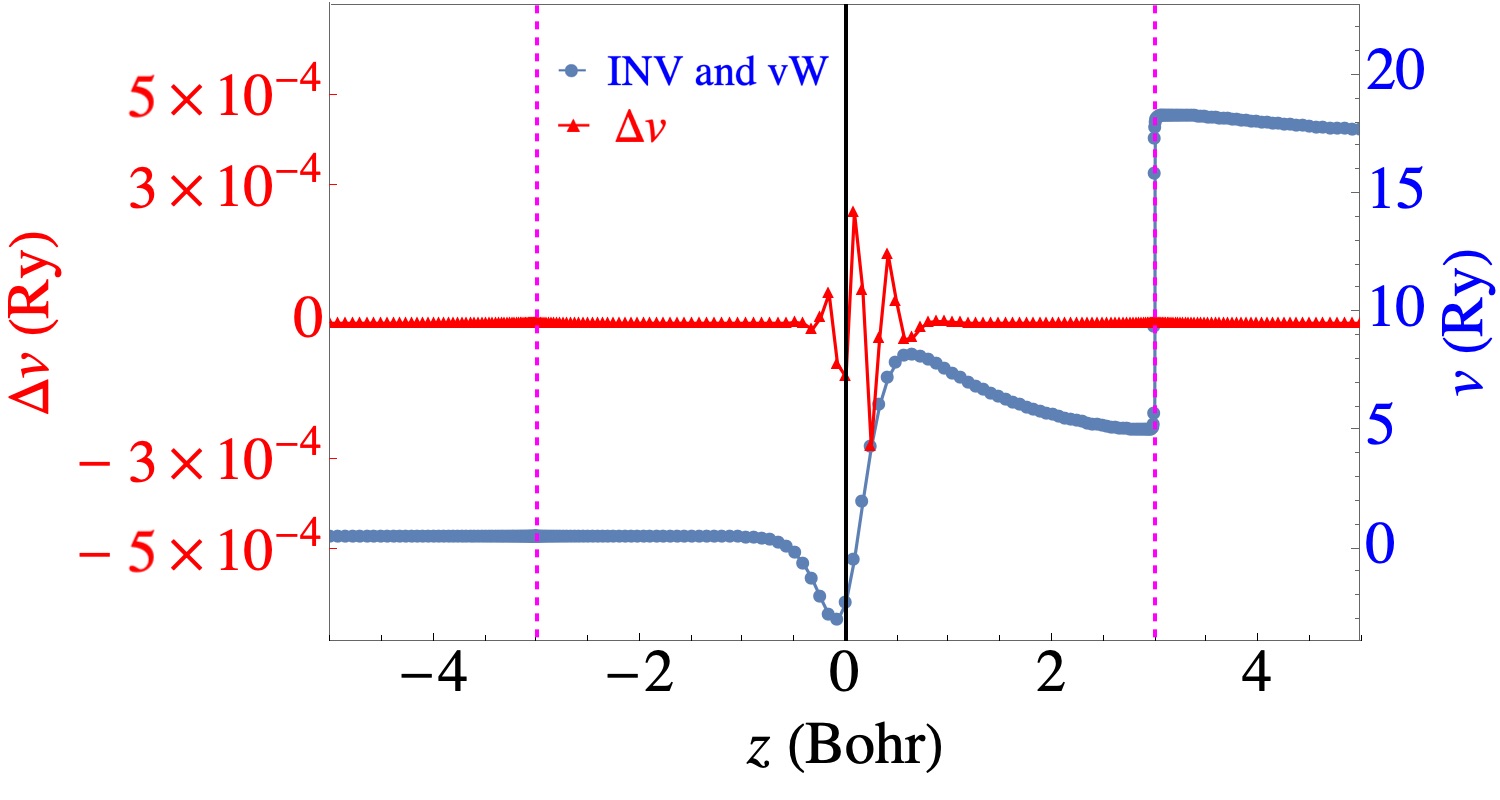}};
    \draw [anchor=north west] (-0.15\linewidth, .95\linewidth) node {\large(a) {\fontfamily{Arial}\selectfont {}}};
    \draw [anchor=north west] (-0.15\linewidth, .25\linewidth) node {\large(b) {\fontfamily{Arial}\selectfont {}}};
\end{tikzpicture}
 \caption{
Non-additive kinetic potential for H$_2$, in a 1D representation, where one electron is localized. Purple dashed lines mark the location of the nuclei and a black solid line marks the cutoff $z_0$.
 (a) Total and partitioned densities.
 (b) Analytically inverted kinetic potential $v^{\rm NAD/INV}[\rho_{\rm B},\rho_{\rm tot}](\textbf{r})$ (from Eq. \eqref{vNADfunctl}), where the localized density $\rho_{\rm B}$ is on the left, compared to the same from von Weizs{\"a}cker theory (from Eq. \eqref{vWvNADfunctl}). Blue: $v^{\rm NAD/INV}$; Red curve: $\Delta v=v^{\text{NAD/vW}} - v^{\rm NAD/INV}$.} 
  \label{1dH2chgDensityH2Invandvw}
 \end{figure}

\begin{figure}[h]
\begin{tikzpicture}

	\node [anchor=north west] (imgA) at (-0.21\linewidth,.90\linewidth)
			{\includegraphics[width=0.95\linewidth]{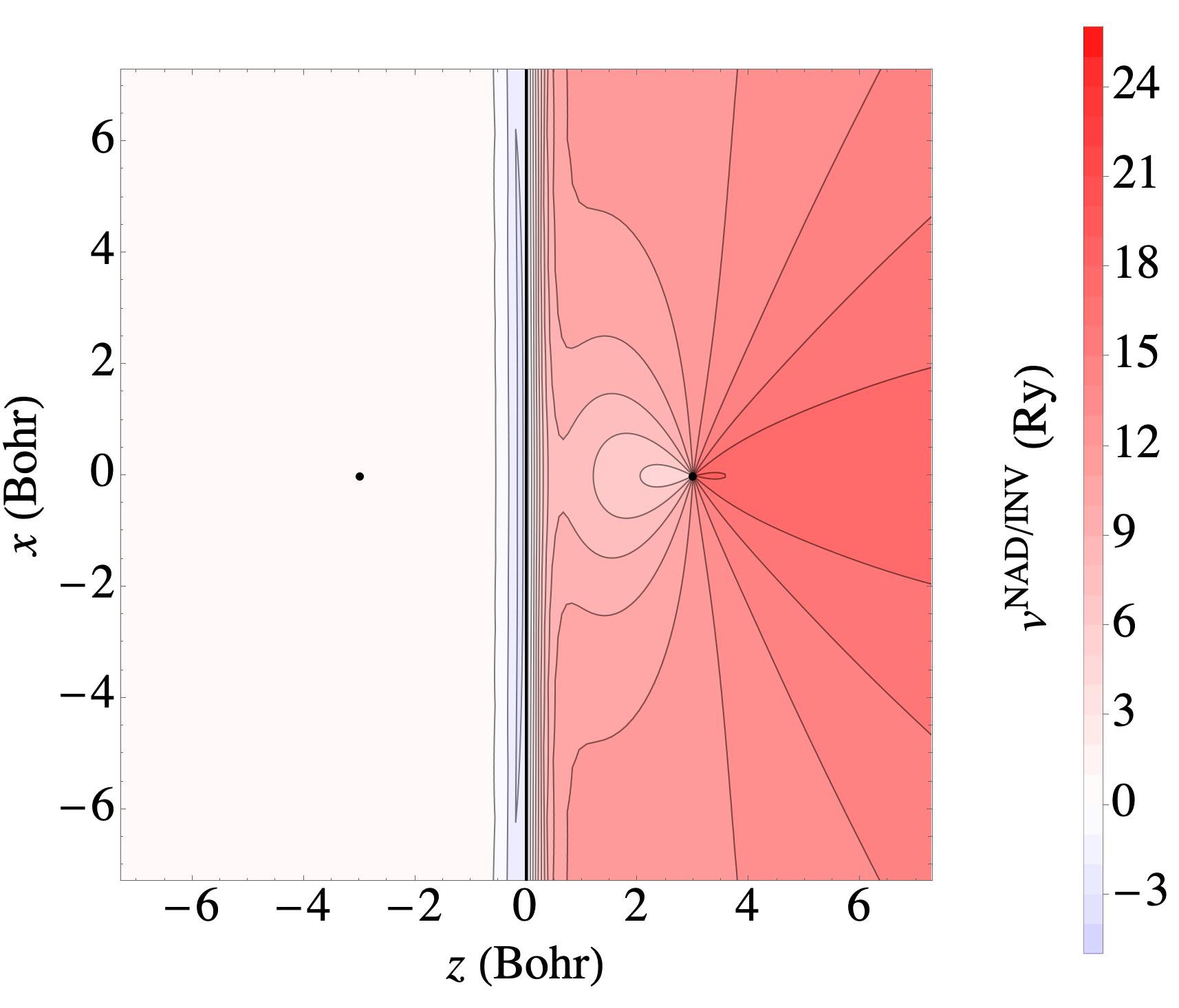}};
   \node [anchor=north west] (imgC) at (-0.21\linewidth,.004\linewidth)
            {\includegraphics[width=0.95\linewidth]{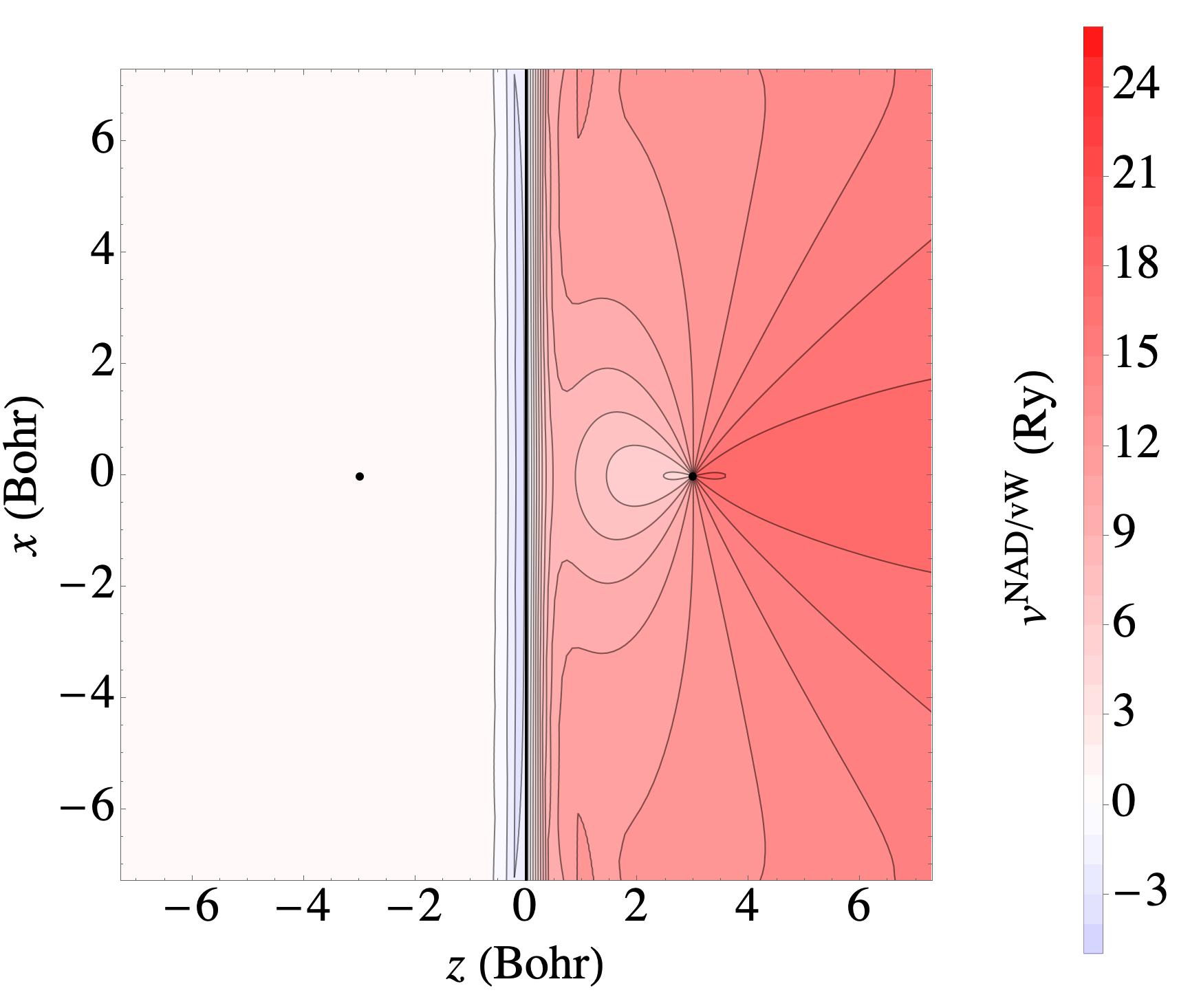}};
    \draw [anchor=north west] (-0.1\linewidth, .83\linewidth) node {\large(a) {\fontfamily{Arial}\selectfont {}}};
    \draw [anchor=north west] (-0.1\linewidth, -.07\linewidth) node {\large(b) {\fontfamily{Arial}\selectfont {}}};
\end{tikzpicture}
 \caption{
 Non-additive kinetic potential for H$_2$ in a 2D representation, comparing (a) $v^{\text{NAD/INV}}$ with (b) $v^{\text{NAD/vW}}$, both in Ry. Black dots mark the nuclei, and a black line marks the cutoff $z_0$.}
 \label{2DH2vW2DH2VNAD} 
 \end{figure}

Our third test system is the stretched H$_2$ system in which one electron is localized around the left nucleus. The cutoff is $z_0 = 0$ by symmetry. The $v^{\rm NAD}$ (Fig. \ref{1dH2chgDensityH2Invandvw}) shows magnitude and features similar to those of HeHe and HeLi$^+$. There is a small attractive well at the overlap region, and then a step and a small well at 5 Ry just to the left of the right nucleus. $v^{\text{NAD/vW}}$ and $v^{\rm NAD/INV}$ agree well with differences again mainly in the overlap region. The 2D view of $v^{\rm NAD}$ (Fig. \ref{2DH2vW2DH2VNAD}) is very similar to HeHe, but with little variation in the attractive well away from the $z$-axis. The only differences between $v^{\rm NAD/INV}$ and $v^{\rm NAD/vW}$ are small deviations in $x>4$ Bohr, in a larger region than for HeHe or HeLi$^+$. No cusps or singularities are seen in the 1D or 2D plots.

\vspace{1.0mm} 

\vspace{1.0mm} 

\section{Conclusion}

In this work, we have investigated analytically and numerically the presence of singularities in the non-additive kinetic potential $v^{\rm NAD}$, and resolved the uncertainty that has surrounded this question, for a variety of important cases of relevance to partitioning schemes.
For two partitioned sub-densities that overlap smoothly and have cusps at all nuclei (Sec. \ref{ClassPaiDensi}), the inverted potential of each has singularities at the vicinity of both nuclei, which then cancel. $v^{\rm NAD}[\rho_{\rm B},\rho_{\rm tot}](\textbf{r})$ has no singularities at the nuclei, and must be smooth everywhere. This situation is convenient for approximations since the singularities are difficult to capture, and a smooth $v^{\rm NAD}$ is easier to use in practical calculations. By contrast, in the usual embedding-theory situation where the sub-density is zero at some nuclei and lacks cusps there (Sec. \ref{sec:embed}), there are singularities in $v^{\rm NAD}$. Singularities can also arise due to the use of Slater-type orbitals or other basis sets that do not perfectly describe cusps (Sec. \ref{sec:finitebasis}). Our analysis applies to both the exact KS potential under some physically likely assumptions; and to all approximated forms known to the authors.

We confirmed these analytic results and demonstrate that they are achievable numerically with the LDA by performing grid-based all-electron calculations for diatomic systems HeHe, HeLi$^+$, and H$_2$, and by computing $v^{\rm NAD}[\rho_{\rm B},\rho_{\rm tot}]$ from analytic inversion. Examination of $v^{\rm NAD}[\rho_{\rm B},\rho_{\rm tot}]$ in 1D and 2D demonstrated smooth behavior and no cusps or singularities, unlike in Ref. \cite{Banafsheh}. We established that the analytically inverted $v^{\rm inv}$ from an orbital and eigenvalue agrees closely with the Kohn-Sham potential used to evaluate the density. We showed a way of calculating $v^{\text{NAD}}[\rho_{\rm B},\rho_{\rm tot}]$ exactly using the von Weizs\"{a}cker potential for these systems, since it is exact for one orbital. We benchmarked our calculations from analytic inversion against $v^{\text{NAD/vW}}[\rho_{\rm B},\rho_{\rm tot}]$ and confirmed the close equality of these potentials throughout space for all 3 diatomic systems, ensuring that the analytically inverted potential is numerically precise and does not suffer from numerical artefacts.

Using our reliable results for $v^{\rm NAD}[\rho_{\rm B},\rho_{\rm tot}]$, we were able to learn about its exact features (for LDA). The potentials shown in Figs. \ref{1dHeHechgDensInvvW}, \ref{1dLiDensDelvInvandvw}, and \ref{1dH2chgDensityH2Invandvw}(b) feature a step and a barrier between the two nuclei.
These are desired and expected features of $v^{\text{NAD}}[\rho_{\rm B},\rho_{\rm tot}](\textbf{r})$ evaluated for a pair of densities $\rho_{\rm B}(\textbf{r})$ and $\rho_{\rm tot}(\textbf{r})$ for which $\rho_{\rm tot}-\rho_{\rm B}$ disappears near the nucleus B.
From a practical perspective, any approximation to  $v^{\text{NAD}}$ should reproduce these features.
Practical calculations using an approximated $v^{\text{NAD}}$ that does not reproduce these features are prone to an artificial leak of electrons onto the nucleus B (for a detailed discussion of this issue see \cite{lastra2008orbital}).

This work has demonstrated exact features of $v^{\rm NAD}$ which can be used to develop and test numerical inversion schemes, kinetic-energy functionals, or non-decomposable approximations that target $v^{\rm NAD}$ directly. The analytic inversion approach can be used for further exploration of exact properties of $v^{\rm NAD}$ in other systems.

\begin{acknowledgments}
We thank Dr. Rachel Garrick for 
help with development in DARSEC,
and Elsa Vazquez for help in preparing plots.
M.B. and D.S. were supported as part of the Consortium for High Energy Density Science by the U.S. Department of Energy, National Nuclear Security Administration, Minority Serving Institution Partnership Program, under Awards DE-NA0003866 and DE-NA0003984, and by UC Merced start-up funds. T.G. was supported by Australian Research Council grants DP200100033 and FT210100663. L.K. thanks the Aryeh and Mintzi Katzman Professorial Chair and the Helen and Martin Kimmel Award for Innovative Investigation. Computational resources were provided by the Multi-Environment Computer for Exploration and Discovery (MERCED) cluster at UC Merced, funded by National Science Foundation Grant No. ACI-1429783.
\end{acknowledgments}

\appendix

\section{Smooth densities, cusps, and non-singular potentials}
\label{App:Smooth}

Densities of electronic systems are finite,
meaning that at any point $\vR_N$ we can make
a series expansion in small $\vr_N=\vr-\vR_N$.
The general formula for expansion of $\rho$ is,
\begin{align}
\rho(\vr)=&\rho_{0,N}
+ b_{\rho}r_N
+ \vec{B}_{\rho}\cdot\vr_N
\nonumber\\&
+ r_N\vec{C}_{\rho}'\cdot\vr_N
+ \vr_N\cdot \mathbf{C}_{\rho}\cdot\vr_N
+ \ldots
\end{align}
where $b_{\rho}$ is a scalar,
$\vec{B}_{\rho}$ and $\vec{C}'_{\rho}$ are vectors and
$\mathbf{C}_{\rho}$ is a $3\times 3$ matrix. This includes
analytic and non-analytic terms.
We may rewrite this as,
\begin{align}
\rho(\vr)=&\rho_{0,N}e^{-2Zr_N} + \rho_{\text{smooth}}(\vr)
\\
\rho_{\text{smooth}}(\vr)=&
\vec{B}_{\rho}\cdot\vr_N
+ r_N\vec{C}_{\rho}'\cdot\vr_N
+ \vr_N\cdot\mathbf{C}''_{\rho}\cdot\vr_N
+ \ldots
\label{eqn:smooth}
\end{align}
where $Z=-b_{\rho}/[2\rho_{0,N}]$ and
$\mathbf{C}''_{\rho}=\mathbf{C}_{\rho}-2Z^2\mathbf{I}$.
Here, we focused on a single nucleus -- the
smooth density must have a similar expansion near
every nucleus.

We therefore obtain,
\begin{align}
\frac{\nabla^2\rho(\vr)}{\rho(\vr)}
=&\frac{-4Z}{r_N}
+ \frac{2\vec{C}'_{\rho}}{\rho_{0,N}}\cdot\hat{\vr}_N
+ \text{const}
\end{align}
where $\hat{\vr}$ indicates a unit vector.
Clearly, the first term dominates, and the
second has a radial average of zero.
This is how the cusp gives rise to
singularities.

The non-singular part, $v_{\text{non-sing}}$,
of the potential obeys,
\begin{align}
\lim_{\vr\to\vR_N}r_N v_{\text{non-sing}}(\vr)=&0,~~~~
\forall \vR_N
\label{vnonsing}
\end{align}
and contains similar terms to Eq.~\eqref{eqn:smooth}.
It can also have a constant
term, a term $B_vr_N$, and even logarithmic
singularities like $L_v\log(r_N)$.
Any non-singular potential obeying Eq.~\eqref{vnonsing}
will not alter any of the conclusions of the main text.

\section{Cusps lead to singularities}
\label{MoreTwo}

For two or few electrons it is trivial to show that singularities
lead to cusps. The leading terms of our density may be
described using $\rho:=e^{-2Z_Nr_N}$,
where $\textbf{r}_N=\textbf{r}-\textbf{R}_N$ is the distance from the cusp
at $\textbf{R}_N$. The remaining terms begin (by definition)
at $O(\textbf{r}_N)$ and therefore contribute to the potential
only at a constant or higher terms.

The non-trivial part of the
von Weizs{\"a}cker potential is therefore the part involving only radial derivatives,
\begin{eqnarray}
\begin{aligned}
v^{\rm vW}=&\frac{\partial_{rr}\rho}{4\rho} + \frac{\partial_r\rho}{2r\rho}
- \frac{(\partial_r\rho)^2}{8\rho^2}
=\frac{Z_N^2}{2} - \frac{Z_N}{r}
\label{eqn:appbvw}
\end{aligned}
\end{eqnarray}
which is clearly dominated by the $\tfrac{-Z_N}{r_N}$
singularity.
This gives our proof for two electrons. To go beyond two
electrons, we show that $v^{\rm vW}$ has the same
singularities as the KS potential for more than two
electrons.

To begin, rewrite $\rho=\sum_i f_i |\phi_i|^2$,
(Eq.~\eqref{eqn:rhoKS}) using,
\begin{eqnarray}
\begin{aligned}
\zeta_i(\textbf{r})\sqrt{\rho(\textbf{r})}:=
\sqrt{f_i}\phi_i(\textbf{r})\;,
\end{aligned}
\end{eqnarray}
where $\sum_{i}|\zeta_i(\textbf{r})|^2=1\forall\vr$,
by definition. The KS orbital
equations Eq.~\eqref{KS-Eq} yield
$[\hat{h}-\epsilon_i]\phi_i=0$, giving,
\begin{eqnarray}
\begin{aligned}
0=&\tfrac{\sqrt{f_i}}{\sqrt{\rho}}
[\tfrac{-\nabla^2}{2} + v_{\rm KS}[\rho] - \epsilon_i]\phi_i
\\
=&\tfrac{1}{\sqrt{\rho}}[\tfrac{-\nabla^2}{2} + v_{\rm KS}[\rho] - \epsilon_i]\zeta_i\sqrt{\rho}
\\
=&[\tfrac{-\nabla^2}{2} - \vec{g}[\rho]\cdot \nabla + \tilde{v}[\rho] - \delta_i]\zeta_i(\textbf{r})
\;,
\label{eqn:xii}
\end{aligned}
\end{eqnarray}
for $\zeta_i$. Here, $\tilde{v}:=v_{\rm KS}[\rho] - v^{\rm vW}[\rho]$,
$\vec{g}[\rho]:=\tfrac{\nabla \sqrt{\rho}}{\sqrt{\rho}}
=\tfrac{\nabla\rho}{2\rho}$, and $\delta_i$ are constants.

Importantly, we recognize that $|\vec{g}|<\infty$
and is smooth in any nuclear density. Since $\delta_i$
is a constant, it follows that only differences
in the location or magnitude of singularities in $v_{\rm KS}$ and
$v^{\rm vW}$, which manifest in $\tilde{v}$, can contribute
to cusps in $\zeta_i$.
We denote the set of effective ``nuclei'' for which
$\tilde{v}$ has singularities by
$\tilde{\FC}=\{(\vR_N,z_N)\}$,
with locations $\vR_N$ and effective charges
$z_N\neq 0\forall N\in\tilde{\FC}$
(we allow $z_N<0$ for repulsive effective nuclei).
Only these singularities may lead to
cusps in $\zeta_i$.

We next recognize that all singularities in $\tilde{v}$
give rise to cusps or zero solutions in $\zeta_i$, which
follows from the series expansion 
of Eq.~\eqref{eqn:xii}
on $r_N=|\vr-\vR_N|$ (see discussion in previous appendix).
Thus, to leading orders, we may define a set
$\zeta_{i\in\vec{I}_c}\approx\zeta_{i,0}[1-z_Nr_N]$
with cusps and a complementary set
$\zeta_{i\notin\vec{I}_c}\approx0$ without cusps,
that are zero at the nuclei.
However, by construction we find $1=\sum_i|\zeta_i|^2
\approx\sum_{i\in\vec{I}_c}|\zeta_{i,0}|^2
-2z_Nr_N\sum_{i\in\vec{I}_c}|\zeta_{i,0}|^2
=C_N - 2z_Nr_N C_N$
where $C_N=\sum_{i\in\vec{I}_N}|\zeta_{i,0}|^2$.
This equation can only be simultaneously correct
for leading and sub-leading order terms
if $z_N=0$, i.e., if there is no singularity
in the vicinity of nucleus $\vR_N$.

Since $z_N=0$ for all nuclei $N$,
it follows that $\tilde{\FC}$ must be the empty set and
that $\tilde{v}$ has no singularities.
The KS potential, $v_{\rm KS}$, therefore
has the same singularities as the von Weizs{\"a}cker
potential, $v^{\rm vW}$, per Eq.~\eqref{eqn:appbvw}.
This extends results to more
than two electrons and yields
\begin{eqnarray}
\begin{aligned}
\sum_N \rho_{0,N}e^{-2Z_Nr_N}
\longrightarrow
\sum_N\frac{-Z_N}{|\textbf{r}-\textbf{R}_N|}
\end{aligned}
\end{eqnarray}
\newline

\end{document}